\documentclass[journal,11pt,draftcls,onecolumn]{IEEEtran}

\usepackage{graphicx}
\usepackage{amsmath, amsthm, amssymb, amsfonts, cancel}
\usepackage{algorithm, algorithmic, ifsym, subfigure}
\usepackage{epstopdf}
\usepackage{siunitx}

\def\tr{{\rm tr}}
\def\E{{\mathbb E}}
\def\P{{\mathbb P}}
\def\bLambda{\boldsymbol{\Lambda}}
\def\bPsi{\boldsymbol{\Psi}}
\def\bXi{\boldsymbol{\Xi}}
\def\bA{ {\mathbf{A}} }
\def\bB{ {\mathbf{B}} }
\def\bC{ {\mathbf{C}} }
\def\bD{ {\mathbf{D}} }
\def\bE{ {\mathbf{E}} }
\def\bF{ {\mathbf{F}} }

\def\bI{ {\mathbf{I}} }

\def\bK{ {\mathbf{K}} }

\def\bM{ {\mathbf{M}} }
\def\bN{ {\mathbf{N}} }

\def\bP{ {\mathbf{P}} }
\def\bQ{ {\mathbf{Q}} }
\def\bR{ {\mathbf{R}} }
\def\bS{ {\mathbf{S}} }

\def\bU{ {\mathbf{U}} }
\def\bV{ {\mathbf{V}} }
\def\bW{ {\mathbf{W}} }
\def\bX{ {\mathbf{X}} }
\def\bY{ {\mathbf{Y}} }
\def\bZ{ {\mathbf{Z}} }
\def\bbeta{ \boldsymbol{\beta} }

\def\bSigma{ {\mathbf{\Sigma}} }
\def\bDelta{ {\mathbf{\Delta}} }
\def\Epsilon{ \mathcal{E} }
\def\bXi{ {\mathbf{\Xi}} }

\def\nn{{ \parallel   }}
\def\RR{{ \mathbb{R}  }}
\def\PP{{ \mathbb{P}  }}
\def\NN{{ \mathbb{N}  }}
\def\Nul{ \si{Nul} }
\def\Col{{ \si{Col} }}
\def\vec{{ \si{vec} }}
\def\tr{{ \si{tr}   }}

\def\diagg{{ \si{diag}   }}
\def\spann{{ \si{span}   }}

\def\rank{ \si{rank}   }

\def\ba{{ \mathbf{a}  }}
\def\bd{{ \mathbf{d}  }}
\def\bx{{ \mathbf{x}  }}
\def\by{{ \mathbf{y}  }}
\def\bu{{ \mathbf{u}  }}
\def\bv{{ \mathbf{v}  }}
\def\bz{{ \mathbf{z}  }}

\def\bv{{ \mathbf{v}  }}
\def\card{{ \text{card} }}

\newtheorem{theorem}{Theorem}
\newtheorem{lemma}{Lemma}

\begin{document}
\title{Covariance Estimation in High Dimensions via Kronecker Product Expansions}
\author{Theodoros Tsiligkaridis *, \textit{Student Member, IEEE}, Alfred O. Hero III, \textit{Fellow, IEEE}}

\maketitle

\begin{abstract}
This paper presents a new method for estimating high dimensional covariance matrices. The method, permuted rank-penalized least-squares (PRLS), is based on a Kronecker product series expansion of the true covariance matrix. Assuming an i.i.d. Gaussian random sample, we establish high dimensional rates of convergence to the true covariance as both the number of samples and the number of variables go to infinity. For covariance matrices of low separation rank, our results establish that PRLS has significantly faster convergence than the standard sample covariance matrix (SCM) estimator. The convergence rate captures a fundamental tradeoff between estimation error and approximation error, thus providing a scalable covariance estimation framework in terms of separation rank, similar to low rank approximation of covariance matrices \cite{Lounici}. The MSE convergence rates generalize the high dimensional rates recently obtained for the ML Flip-flop algorithm \cite{KGlasso13, TsiligkaridisTSP} for Kronecker product covariance estimation. We show that a class of block Toeplitz covariance matrices is approximatable by low separation rank and give bounds on the minimal separation rank $r$ that ensures a given level of bias. Simulations are presented to validate the theoretical bounds. As a real world application, we illustrate the utility of the proposed Kronecker covariance estimator for spatio-temporal linear least squares prediction of multivariate wind speed measurements.
\end{abstract}

\begin{keywords}
	Structured covariance estimation, penalized least squares, Kronecker product decompositions, high dimensional convergence rates, mean-square error, multivariate prediction.
\end{keywords}

\let\thefootnote\relax\footnote{The research reported in this paper was supported in part by ARO grant W911NF-11-1-0391. Preliminary results in this paper have appeared at the 2013 IEEE International Symposium on Information Theory.

T. Tsiligkaridis and A. O. Hero, III, are with the Department of Electrical Engineering and Computer Science, University of Michigan, Ann Arbor, MI 48109 USA (e-mail: ttsili@umich.edu, hero@umich.edu).
}

\section{Introduction}
Covariance estimation is a fundamental problem in multivariate statistical analysis. It has received attention in diverse fields including economics and financial time series analysis (e.g., portfolio selection, risk management and asset pricing \cite{Bai2011}), bioinformatics (e.g. gene microarray data \cite{Gene2003, HeroRaj2012}, functional MRI \cite{Biometrics2010}) and machine learning (e.g., face recognition \cite{ZhangNIPS10}, recommendation systems \cite{AllenTib10}). In many modern applications, data sets are very large with both large number of samples $n$ and large dimension $d$, often with $d\gg n$, leading to a number of covariance parameters that greatly exceeds the number of observations. The search for good low-dimensional representations of these data sets has led to much progress in their analysis. Recent examples include sparse covariance estimation \cite{YL07, ModelSel, RWRY08, Rothman}, low rank covariance estimation \cite{Fan2008, Fitz2004, Johnstone2009, Lounici}, and Kronecker product esimation \cite{Dutilleul, EstCovMatKron, Dawid, KGlasso13, TsiligkaridisTSP}. 

%

Kronecker product (KP) structure is a different covariance constraint from sparse or low rank constraints. KP represents a $pq \times pq$ covariance matrix $\bSigma_0$ as the Kronecker product of two lower dimensional covariance matrices. When the variables are multivariate Gaussian with covariance following the KP model, the variables are said to follow a matrix normal distribution \cite{Dawid, Dutilleul, GuptaNagar99}. This model has applications in channel modeling for MIMO wireless communications \cite{MIMOWerner}, geostatistics \cite{Cressie93}, genomics \cite{YinLi2012},  multi-task learning \cite{BonillaNIPS08}, face recognition \cite{ZhangNIPS10}, recommendation systems  \cite{AllenTib10} and collaborative filtering \cite{YuICML09}. 
The main difficulty in maximum likelihood estimation of structured covariances is the nonconvex optimization problem that arises. Thus, an alternating optimization approach is usually adopted. In the case where there is no missing data, an extension of the alternating optimization algorithm of Werner {\it et al} \cite{EstCovMatKron}, that the authors called the flip flop (FF) algorithm, can be applied to estimate the parameters of the Kronecker product model, called KGlasso in \cite{KGlasso13}. 

In this paper, we assume that the covariance can be represented as a sum of Kronecker products of two lower dimensional factor matrices, where the number of terms in the summation may depend on the factor dimensions. More concretely, we assume that there are $d=pq$ variables whose covariance $\mathbf{\Sigma}_0$ has Kronecker product representation:
\begin{equation} \label{eq: KP_model}
	\mathbf{\Sigma}_{0} = \sum_{\gamma=1}^r \mathbf{A}_{0,\gamma} \otimes \mathbf{B}_{0,\gamma}
\end{equation}
where $\{\bA_{0,\gamma}\}$ are $p\times p$ linearly independent matrices and $\{\bB_{0,\gamma}\}$ are $q \times q$ linearly independent matrices \footnote{Linear independence is with respect to the trace inner product defined in the space of symmetric matrices.}. We assume that the factor dimensions $p,q$ are known. We note $1\leq r\leq r_0=\min(p^2,q^2)$ and refer to $r$ as the {\it separation rank}. The model (\ref{eq: KP_model}) is analogous to separable approximation of continuous functions \cite{Beylkin:2005}. It is evocative of a type of low rank principal component decomposition where the components are Kronecker products. However, the components in (\ref{eq: KP_model}) are neither orthogonal nor normalized. The model (\ref{eq: KP_model}) with separation rank 1 is relevant to channel modeling for MIMO wireless communications, where $\bA_0$ is a transmit covariance matrix and $\bB_0$ is a receive covariance matrix \cite{MIMOWerner}. The rank 1 model is also relevant to other transposable models arising in recommendation systems like NetFlix and in gene expression analysis \cite{AllenTib10}. The model (\ref{eq: KP_model}) with $r\geq 1$ has applications in spatiotemporal MEG/EEG covariance modeling \cite{deMunck2002, deMunck2004, Bijma05, Jun2006} and SAR data analysis \cite{SAR2010}. We finally note that Van Loan and Pitsianis \cite{VanLoanKP} have shown that any $pq\times pq$ matrix $\bSigma_0$ can be written as an orthogonal expansion of Kronecker products of the form (\ref{eq: KP_model}), thus allowing any covariance matrix to be approximated by a bilinear decomposition of this form.

The main contribution of this paper is a convex optimization approach to estimating covariance matrices with KP structure of the form (\ref{eq: KP_model}) and the derivation of tight high-dimensional MSE convergence rates as $n$, $p$ and $q$ go to infinity. We call our method the Permuted Rank-penalized Least Squares (PRLS) estimator. Similarly to other studies of high dimensional covariance estimation \cite{CaiZhangZhou:2010, KGlasso13, Rothman, BickelLevina:2008, Vershynin}, we analyze the estimator convergence rate in Frobenius norm of PRLS, providing specific convergence rates holding with certain high probability. In other words, our anlaysis provides high probability guarantees up to absolute constants in all sample sizes and dimensions.

For estimating separation rank $r$ covariance matrices of the form (\ref{eq: KP_model}), we establish that PRLS achieves high dimensional consistency with a convergence rate of $O_P\left(\frac{r (p^2+q^2+\log \max(p,q,n))}{n}\right)$. This can be significantly faster than the convergence rate $O_P\left(\frac{p^2 q^2}{n}\right)$ of the standard sample covariance matrix (SCM). For separation rank $r=1$ this rate is identical to that of the FF algorithm, which fits the sample covariance matrix to a single Kronecker factor.

The PRLS method for estimating the Kronecker product expansion (\ref{eq: KP_model}) generalizes previously proposed Kronecker product covariance models \cite{Dutilleul, Dawid} to the case of $r>1$. This is a fundamentally different generalization than the $r=1$ sparse KP models proposed in \cite{AllenTib10, KGlasso13, TsiligkaridisTSP, LengTang2012}. Independently in \cite{KGlasso13, TsiligkaridisTSP} and \cite{LengTang2012}, it was established that the high dimensional convergence rate for these sparse KP models is of order $O_P\left(\frac{(p+q) \log \max(p,q,n)}{n} \right)$. While we do not pursue the the additional constraint of sparsity in this paper, we speculate that sparsity can be combined with the Kronecker sum model (\ref{eq: KP_model}), achieving even better convergence.

Advantages of the proposed PRLS covariance estimator is illustrated on both simulated and real data. The application of PRLS to the NCEP wind dataset shows that a low order Kronecker sum provides a remarkably good fit to the spatio-temporal sample covariance matrix: over $86 \%$ of all the energy is contained in the first Kronecker component of the Kronecker expansion as compared to only $41 \%$ in the principal component of the standard PCA eigen-expansion. Furthermore, by replacing the SCM in the standard linear predictor by our Kronecker sum estimator we demonstrate a $1.9$ dB RMSE advantage for predicting next-day wind speeds from NCEP network past measurements. 

The outline of the paper is as follows. Section \ref{sec: notation} introduces the notation that will be used throughout the paper. Section \ref{sec: prls} introduces the PRLS covariance estimation method. Section \ref{sec: consistency} presents the high-dimensional MSE convergence rate of PRLS. Section \ref{sec: simulations} presents numerical experiments. The technical proofs are placed in the Appendix.

\section{Notation} \label{sec: notation}
For a square matrix $\mathbf{M}$, define $|\mathbf{M}|_1=\nn \vec(\mathbf{M}) \nn_1$ and $|\mathbf{M}|_\infty=\nn \vec(\mathbf{M}) \nn_\infty$, where $\vec(\mathbf{M})$ denotes the vectorized form of $\mathbf{M}$ (concatenation of columns of $\bM$ into a column vector). $\nn \mathbf{M} \nn_2$ is the spectral norm of $\bM$. $\mathbf{M}_{i,j}=[\mathbf{M}]_{i,j}$ is the $(i,j)$th element of $\mathbf{M}$. Let the inverse transformation (from a vector to a matrix) be defined as: $\vec^{-1}(\mathbf{x})=\mathbf{X}$, where $\mathbf{x}=\vec(\mathbf{X})$. Define the $pq\times pq$ permutation operator $\bK_{p,q}$ such that $\bK_{p,q} \vec(\bN) = \vec(\bN^T)$ for any $p\times q$ matrix $\bN$. For a symmetric positive definite matrix $\bM$, $\lambda(\mathbf{M})$ will denote the vector of real eigenvalues of $\mathbf{M}$ and define $\lambda_{max}(\mathbf{M})=\nn\bM \nn_2=\max{\lambda_i(\mathbf{M})}$, and $\lambda_{min}(\mathbf{M}) = \min{\lambda_i(\mathbf{M})}$. 
For any matrix $\bM$, define the nuclear norm $\nn \bM \nn_* = \sum_{l=1}^{r_M} |\sigma_l(\bM)|$, where $r_M = \rank(\bM)$ and $\sigma_l(\bM)$ is the $l$th singular value of $\bM$.

\begin{figure}[ht]
	\centering
		\includegraphics[width=0.70\textwidth]{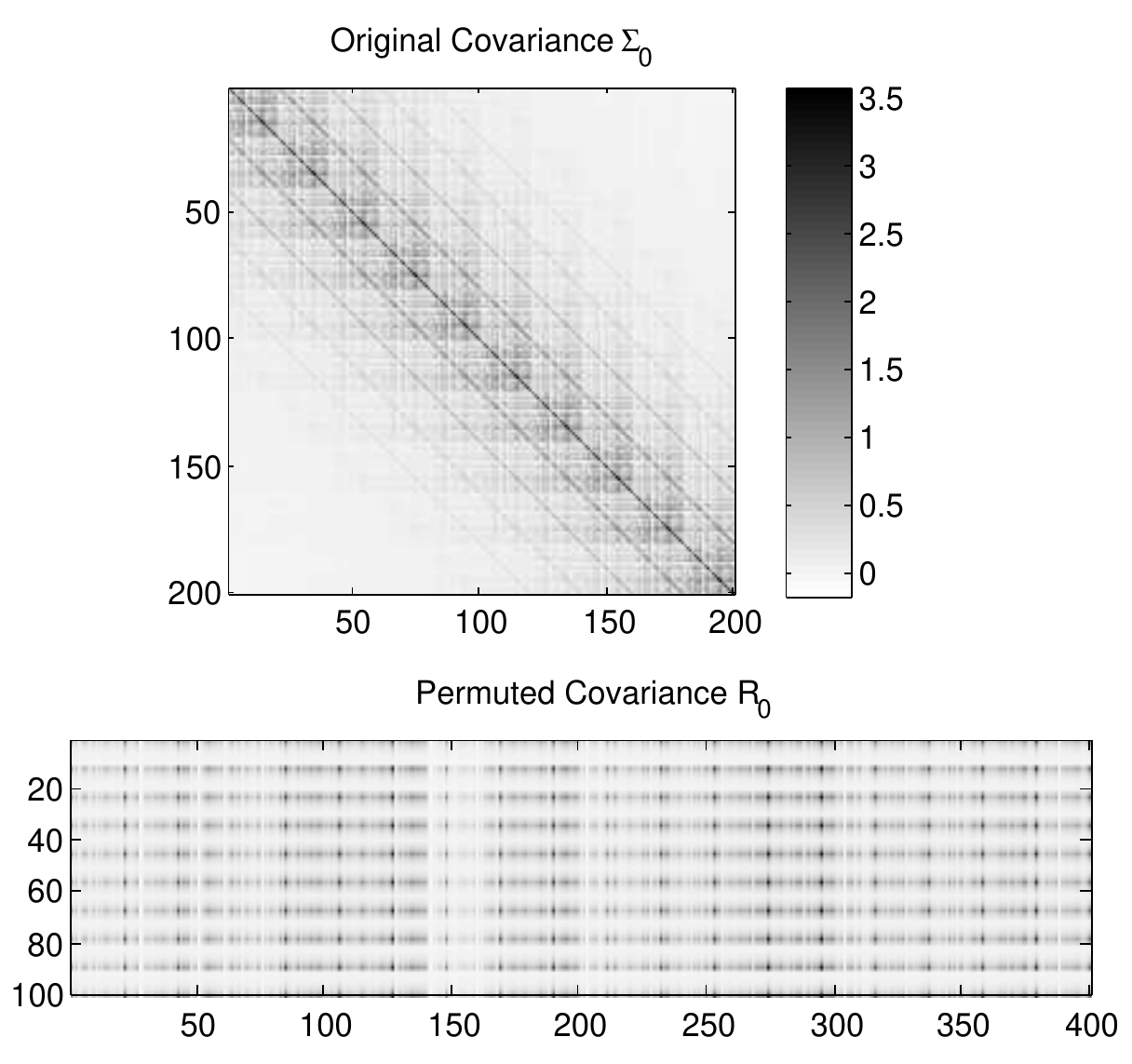}
	\caption{\label{fig:matrices_perm} Original (top) and permuted covariance (bottom) matrix. The original covariance is $\bSigma_0=\bA_0\otimes \bB_0$, where $\bA_0$ is a $10\times 10$ Toeplitz matrix and $\bB_0$ is a $20 \times 20$ unstructured p.d. matrix. Note that the permutation operator $\mathcal{R}$ maps a symmetric p.s.d. matrix $\bSigma_0$ to a non-symmetric rank 1 matrix $\bR_0 = \mathcal{R}(\bSigma_0)$. }
\end{figure}

For a matrix $\bM$ of size $pq\times pq$, let $\{\bM(i,j)\}_{i,j=1}^p$ denote its $q\times q$ block submatrices, where each block submatrix is $\bM(i,j)=[\bM]_{(i-1)q+1:iq,(j-1)q+1:jq}$. Also let $\{\overline{\bM}(k,l)\}_{k,l=1}^q$ denote the $p\times p$ block submatrices of the permuted matrix $\overline{\bM}=\bK_{p,q}^T \bM \bK_{p,q}$. Define the permutation operator $\mathcal{R}: \RR^{pq\times pq} \to \RR^{p^2 \times q^2}$ by setting the $(i-1)p+j$ row of $\mathcal{R}(\bM)$ equal to $\vec(\bM(i,j))^T$. When $\bM$ is representable as the Kronecker product $\bM_1\otimes \bM_2$, an illustration of this permutation operator is shown in Fig. \ref{fig:matrices_perm}.

Define the set of symmetric matrices $S^p = \{\mathbf{A}\in \RR^{p\times p}: \mathbf{A}=\mathbf{A}^T\}$, the set of symmetric positive semidefinite (psd) matrices $S_{+}^p$, and the set of symmetric positive definite (pd) matrices $S_{++}^p$. $\bI_d$ is a $d\times d$ identity matrix. It can be shown that $S_{++}^p$ is a convex set but is not closed \cite{Boyd}. Note that $S_{++}^p$ is simply the interior of the closed convex cone $S_{+}^p$.

For a subspace $U$, define $\bP_U$ and $\bP_U^{\perp}$ as the orthogonal projection operators projecting onto $U$ and $U^{\perp}$, respectively. The unit Euclidean sphere in $\RR^{d'}$ is denoted by $\mathcal{S}_{d-1}=\{\bx \in \RR^d: \nn \bx\nn_2=1\}$. Let $(x)_+=\max(x,0)$.

Statistical convergence rates will be denoted by the $O_P(\cdot)$ notation, which is defined as follows. Consider a sequence of real random variables $\{X_n\}_{n \in \NN}$ defined on a probability space $(\Omega,\mathcal{F},P)$ and a deterministic (positive) sequence of reals $\{b_n\}_{n\in \NN}$. By $X_n=O_P(1)$ is meant: $\sup_{n\in \NN}{\Pr(|X_n|>K)} \to 0$ as $K\to\infty$, where $X_n$ is a sequence indexed by $n$, for fixed $p,q$. 
The notation $X_n=O_P(b_n)$ is equivalent to $\frac{X_n}{b_n}=O_P(1)$. By $X_n=o_p(1)$ is meant $\Pr(|X_n|>\epsilon) \to 0$ as $n\to\infty$ for any $\epsilon>0$. By $\lambda_n \asymp b_n$ is meant $c_1 \leq \frac{\lambda_n}{b_n} \leq c_2$ for all $n$, where $c_1,c_2>0$ are finite  constants.

\section{Permuted Rank-penalized Least-squares} \label{sec: prls}

Available are $n$ i.i.d. multivariate Gaussian observations $\{\bz_t\}_{t=1}^n$, $\bz_t\in \RR^{pq}$, having zero-mean and covariance equal to (\ref{eq: KP_model}). A sufficient statistic for estimating the covariance is the well-known sample covariance matrix (SCM):
\begin{equation} \label{eq:SCM}
	\hat{\bS}_n = \frac{1}{n} \sum_{t=1}^n \bz_t \bz_t^T
\end{equation}
The SCM is an unbiased estimator of the true covariance matrix. However, when the number of samples $n$ is smaller than the number of variables $d=pq$ the SCM suffers from high variance and a low rank approximation to the SCM is commonly used. The most common low rank approximation is to perform the eigendecomposition of $\hat{\bS}_n$ and retain only the top $r$ principal components resulting in an estimator, called the PCA estimator, of the form:
\begin{equation} \label{eqn:PCA}
	\hat{\bS}_n^{PCA} = \sum_{i=1}^r \sigma^2_i  \nu_i\nu_i^T,
\end{equation}
where $r<d$ is selected according to some heuristic. It is now well known \cite{Lee:2010, Rao:2008} that this PCA estimator suffers from high bias when $n$ is smaller than $d=pq$.

An alternative approach to low rank covariance estimation was proposed in \cite{Lounici} specifying a low rank covariance estimator as the solution of the penalized least squares problem \footnote{The estimator (\ref{eq:SVT}) was developed in \cite{Lounici} for the more general problem where there could be missing data.}:
\begin{equation} \label{eq:SVT}
	\hat{\bSigma}_n^\lambda  \in \arg \min_{\bS \in S_{++}^d} \nn \hat{\bS}_n - \bS\nn_F^2 + \lambda \tr(\bS)
\end{equation}
where $\lambda>0$ is a regularization parameter. 

The estimator (\ref{eq:SVT}) has several useful interpretations. First, it can be interpreted as a convex relaxation of the non-convex rank constrained Frobenius norm minimization problem
\begin{equation*}
	\arg\min_{{\bS} \in S_{++}^d, \rank(\bS)\leq r} \parallel \hat{\bS}_n-\bS \parallel_F^2,
\end{equation*}
whose solution, by the Eckhart-Young theorem, is the PCA estimator (\ref{eqn:PCA}). Second, it can be interpreted as a covariance version of the lasso regression problem, i.e., finding a low rank psd $\ell_2$ approximation to the sample covariance matrix. The term $\tr(\bS)$ in \ref{eq:SVT} is equivalent to the $\ell_1$ norm on the eigenvalues of the psd matrix $\mathbf S$. As shown in \cite{Lounici} the solution to the convex minimization in (\ref{eq:SVT}) converges to the ensemble covariance $\bSigma_0 = \E[\mathbf z_t\mathbf z_t^T]$ at the minimax optimal rate. Corollary 1 in \cite{Lounici} establishes that, for $\lambda = C' \nn \bSigma_0\nn_2 \sqrt{\frac{r(\bSigma_0) \log(2d)}{n}}$, $n \geq c r(\bSigma_0) \log^2(\max(2d,n))$ and $C',c>0$ sufficiently large, establishes a tight Frobenius norm error bound, which states that with probability $1-\frac{1}{2d}$:
\begin{equation*}
	\nn \hat{\bSigma}_n^\lambda - \bSigma_0 \nn_F^2 \leq \inf_{\bS \succ 0} \nn \bSigma_0-\bS \nn_F^2 + C \nn \bSigma_0 \nn_2^2 \rank(\bS) \frac{r(\bSigma_0) \log(2d)}{n}
\end{equation*}
where $r(\bSigma_0) = \frac{\tr(\bSigma_0)}{\nn \bSigma_0 \nn_2} \leq \min\{\rank(\bSigma_0), d\}$ is the effective rank \cite{Lounici}. The absolute constant $C$ is given by $\frac{(1+\sqrt{2})^2}{8}(C')^2$.


Here we propose a similar nuclear norm penalization approach to estimate low separation-rank covariance matrices of form (\ref{eq: KP_model}). Motivated by Van Loan and Pitsianis's work \cite{VanLoanKP}, we propose:
\begin{equation} \label{eq: prls_problem}
	\hat{\bR}_n^\lambda \in \arg \min_{\bR \in \RR^{p^2\times q^2}} \nn \hat{\bR}_n - \bR \nn_F^2 + \lambda \nn \bR \nn_*
\end{equation}
where $\hat{\bR}_n= \mathcal{R}(\hat{\bS}_n)$ is the permuted SCM of size $p^2 \times q^2$ (see Notation section). The minimum-norm problem considered in \cite{VanLoanKP} is:
\begin{equation} \label{eq: vanloan}
	\min_{\bR \in \RR^{p^2\times q^2}: \rank(\bR) \leq r} \nn \hat{\bR}_n - \bR \nn_F^2
\end{equation}
Specifically, let $\bS=\sum_{i=1}^r \bA_i\otimes \bB_i$ where for all $i$ the dimensions of the matrices $\bA_i$ and $\bB_i$ are fixed. Then, as the Frobenius norm of a matrix is invariant to permutation of its elements, it follows that $\parallel \bS_n-\bS\parallel_F = \parallel \bR_n-\bR\parallel_F$ where $\bR_n=\mathcal{R}(\bS_n)$ and $\bR=\mathcal{R}(\bS)$ (which is a matrix of algebraic rank $r$).

We note that (\ref{eq: prls_problem}) is a convex relaxation of (\ref{eq: vanloan}) and is more amenable to numerical optimization. Furthermore, we show a tradeoff between approximation error (i.e., the error induced by model mismatch between the true covariance and the model) and estimation error (i.e., the error due to finite sample size) by analyzing the solution of (\ref{eq: prls_problem}). We also note that (\ref{eq: prls_problem}) is a strictly convex problem, so there exists a unique solution that can be efficiently found using well established numerical methods \cite{Boyd}.

The solution of (\ref{eq: prls_problem}) is closed form and is given by a thresholded singular value decomposition:
\begin{equation} \label{eq:solution}
	\hat{\bR}_n^\lambda = \sum_{j=1}^{\min(p^2,q^2)} \left( \sigma_j(\hat{\bR}_n)-\frac{\lambda}{2} \right)_+ \bu_j \bv_j^T
\end{equation}
where $\bu_j$ and $\bv_j$ are the left and right singular vectors of $\hat{\bR}_n$. This is converted back to a square $pq\times pq$ matrix $\hat{\bSigma}_n^\lambda$ by applying the inverse permutation operator $\mathcal{R}^{-1}$ to $\hat{\bR}_n$ (see Notation section).

Efficient methods for numerically evaluating penalized objectives like (\ref{eq: prls_problem}) have been recently proposed \cite{CaiCandesShen2010, CaiOsher2010} and do not require computing the full SVD. Although empirically observed to be fast, the computational complexity of the algorithms presented in \cite{CaiCandesShen2010} and \cite{CaiOsher2010} is unknown. The rank-$r$ SVD can be computed with $O(p^2 q^2 r)$ floating point operations. There exist faster randomized methods for truncated SVD requiring only $O(p^2 q^2 \log(r))$ floating point operations \cite{HMT2010}. Thus, the computational complexity of solving (\ref{eq: prls_problem}) scales well with respect to the desired separation rank $r$.

The next theorem shows that the de-permuted version of (\ref{eq:solution}) is symmetric and positive definite.
\begin{theorem} \label{thm: symmetry_pd}
Consider the de-permuted solution $\hat{\bSigma}_n^\lambda=\mathcal{R}^{-1}(\hat{\bR}_n^\lambda)$. The following are true:
	\begin{enumerate}
		\item	The solution $\hat{\bSigma}_n^\lambda$ is symmetric with probability 1.
		\item If $n\geq pq$, then the solution $\hat{\bSigma}_n^\lambda$ is positive definite with probability 1.
	\end{enumerate}
\end{theorem}
\begin{IEEEproof}
	See Appendix A.
\end{IEEEproof}
We believe that the PRLS estimate $\hat{\bSigma}_n^\lambda$ is positive definite even if $n<pq$ for appropriately selected $\lambda>0$. In our simulations, we always found $\hat{\bSigma}_n^\lambda$ to be positive definite. We have also found that the condition number of the PRLS estimate is orders of magnitude smaller than that of the SCM.

\section{High Dimensional Consistency of PRLS} \label{sec: consistency}
In this section, we show that RPLS achieves the MSE statistical convergence rate of $O_P \left( \frac{r (p^2 + q^2 + \log M)}{n} \right)$. This result is clearly superior to the statistical convergence rate of the naive SCM estimator \cite{Vershynin},
\begin{equation} \label{eq: SCM_rate}
	\nn \hat{\bS}_n - \bSigma_0 \nn_F^2 = O_P\left( \frac{p^2 q^2}{n} \right),
\end{equation}
particularly when $p,q\to\infty$.

The next result provides a relation between the spectral norm of $\hat{\bR}_n - \bR_0$, the Frobenius norm of $\bR-\bR_0$ and the Frobenius norm of the the estimation error $\hat{\bR}_n^\lambda-\bR_0$.
\begin{theorem} \label{thm: Frob_rate}
	Consider the convex optimization problem (\ref{eq: prls_problem}). When $\lambda\geq 2 \nn \hat{\bR}_n - \bR_0 \nn_2$, the following holds:
	\begin{equation} \label{eq: Frob_rate}
		\nn \hat{\bR}_n^\lambda-\bR_0 \nn_F^2 \leq \inf_{\bR} \left\{ \nn \bR-\bR_0\nn_F^2 + \frac{(1+\sqrt{2})^2}{4} \lambda^2 \rank(\bR) \right\}
	\end{equation}
\end{theorem}
\begin{IEEEproof}
	See Appendix B.
\end{IEEEproof}

\subsection{High Dimensional Operator Norm Bound for the Permuted Sample Covariance Matrix}

In this subsection, we establish a tight bound on the spectral norm of the error matrix
\begin{equation} \label{eq:bDelta_n}
	\bDelta_n = \hat{\bR}_n-\bR_0 = \mathcal{R}(\hat{\bS}_n - \bSigma_0).
\end{equation}
The standard strong law of large numbers implies that for fixed dimensions $p,q$, we have $\bDelta_n \to 0$ almost surely as $n\to\infty$. The next result will characterize the finite sample fluctuations of this convergence (in probability) measured by the spectral norm as a function of the sample size $n$ and Kronecker factor dimensions $p,q$. This result will be useful for establishing a tight bound on the Frobenius norm convergence rate of PRLS and can guide the selection of the regularization paramater in (\ref{eq: prls_problem}).
\begin{theorem} (Operator Norm Bound on Permuted SCM) \label{thm: Operator_rate}
	Assume $\nn \bSigma_0 \nn_2<\infty$ for all $p,q$ and define $M=\max(p,q,n)$. Fix $\epsilon' < \frac{1}{2}$. Assume $t\geq \max(\sqrt{4C_1 \ln(1+\frac{2}{\epsilon'})}, 4C_2 \ln(1+\frac{2}{\epsilon'}))$ and $C=\max(C_1,C_2)>0$ \footnote{The constants $C_1,C_2$ are defined in Lemma \ref{lemma: large_dev} in Appendix B.}. Then, with probability at least $1-2M^{-\frac{t}{4C}}$,
	\begin{equation} \label{eq: Operator_rate}
		\nn \bDelta_n \nn_2 \leq \frac{C_0 t}{1-2\epsilon'} \max\left\{ \frac{p^2 + q^2 + \log M}{n} , \sqrt{\frac{p^2 + q^2 + \log M}{n}} \right\}
	\end{equation}
	where $C_0=\nn \bSigma_0 \nn_2 >0$ \footnote{The constant $\frac{C_0 t}{1-2\epsilon'}$ in front of the rate can be optimized by minimizing it as a function of $\epsilon'$ over the interval $(0,1/2)$.}.
\end{theorem}
\begin{IEEEproof}
	See Appendix D.
\end{IEEEproof}
The proof technique is based on a large deviation inequality, derived in Lemma \ref{lemma: large_dev} in Appendix C. This inequality  characterizes the tail behavior of the quadratic form $\bx^T \bDelta_n \by$ over the spheres $\bx \in \mathcal{S}_{p^2-1}$ and $\by \in \mathcal{S}_{q^2-1}$. Using Lemma \ref{lemma: large_dev} and a sphere covering argument, the result of Theorem \ref{thm: Operator_rate} follows (see Appendix E). Fig. \ref{fig:spec_norm_bnd} empirically validates the tightness of the bound (\ref{eq: Operator_rate}) under the trivial separation rank 1 covariance $\bSigma_0 = \bI_{p} \otimes \bI_{q}$.
\begin{figure}[ht]
	\centering
		\includegraphics[width=0.80\textwidth]{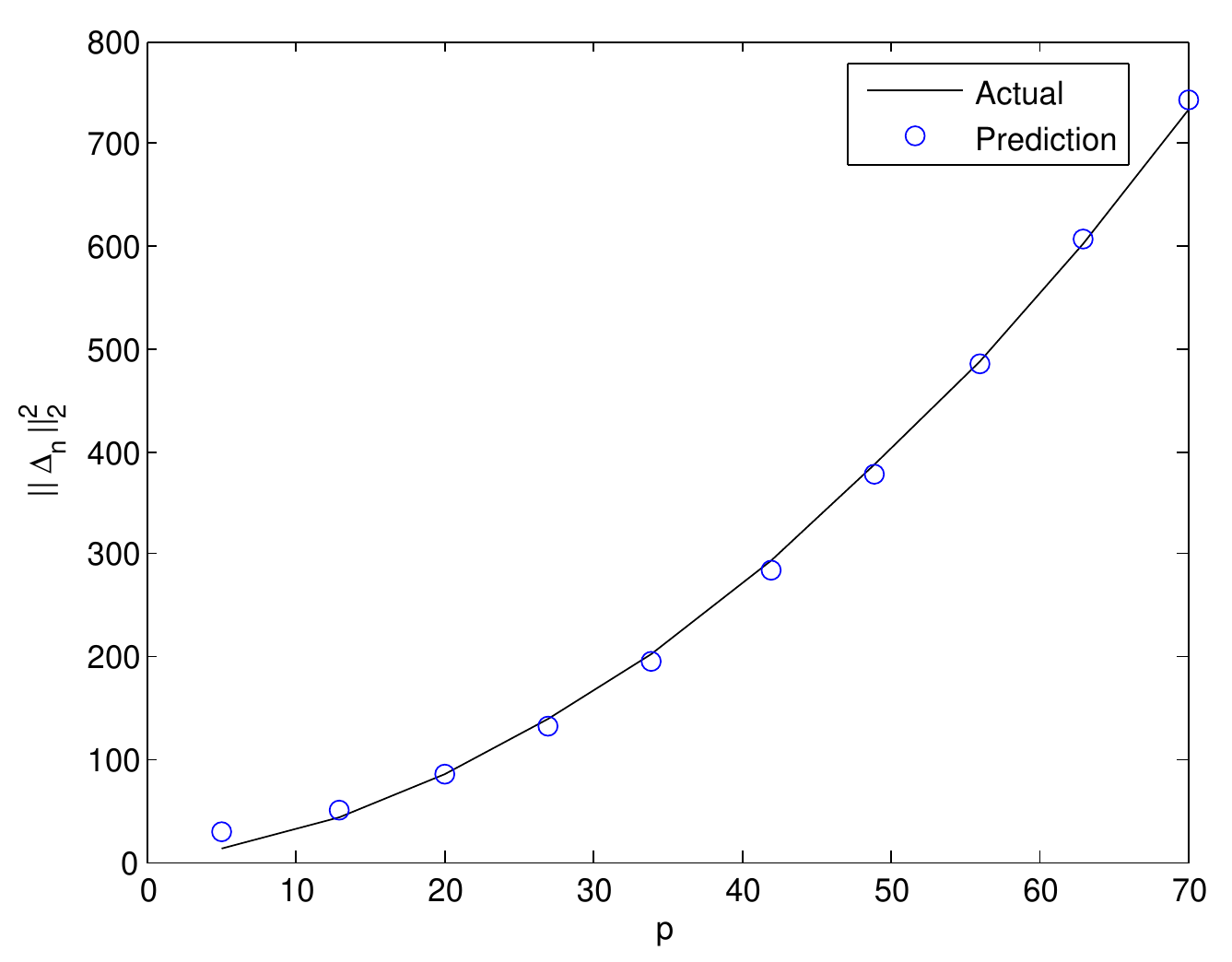}
	\caption{ \label{fig:spec_norm_bnd} Monte Carlo simulation for growth of spectral norm $\nn\bDelta_n\nn_2^2$ as a function of $p$ for fixed $n=10$ and $q=5$. The predicted curve is a least-square fit of a quadratic model $y=ax^2+b$ to the empirical curve, and is a great fit. This example shows the tightness of the probabilistic bound (\ref{eq: Operator_rate}). }
\end{figure}

\subsection{High Dimensional MSE Convergence Rate for PRLS}

Using the result in Thm. \ref{thm: Operator_rate} and the bound in Thm. \ref{thm: Frob_rate}, we next provide a tight bound on the MSE estimation error.
\begin{theorem} \label{thm: PRLS_Frob_rate}
	Define $M=\max(p,q,n)$. Set $\lambda = \lambda_n = \frac{2 C_0 t}{1-2\epsilon'} \max\left\{ \frac{p^2 + q^2 + \log M}{n}, \sqrt{\frac{p^2 + q^2 + \log M}{n}} \right\}$ with $t$ satisfying the conditions of Thm. \ref{thm: Operator_rate}. Then, with probability at least $1-2M^{-\frac{t}{4C}}$:
	\begin{align} 
		\nn & \hat{\bSigma}_n^\lambda - \bSigma_0 \nn_F^2 \leq \inf_{\bR: \rank(\bR) \leq r} \nn \bR-\bR_0 \nn_F^2 \nonumber \\
			&+ C' r \max\left\{ \left(\frac{p^2 + q^2 + \log M}{n}\right)^2 , \frac{p^2 + q^2 + \log M}{n} \right\} \label{eq: PRLS_Frob_rate}
	\end{align}
	where $C'=\left(C_0 t \frac{1+\sqrt{2}}{1-2\epsilon'}\right)^2=\left(3(1+\sqrt{2}) C_0 t\right)^2 > 0$.
\end{theorem}
\begin{IEEEproof}
	See Appendix E.
\end{IEEEproof}

When $\bSigma_0$ is truly a sum of $r$ Kronecker products with factor dimensions $p$ and $q$, there is no model mismatch and the approximation error $\inf_{\left\{\bR: \rank(\bR)\leq r\right\}} \nn \bR-\bR_0 \nn_F^2$ is zero. In this case, in the large-$p,q,n$ asymptotic regime where $p^2 + q^2 + \log M = o(n)$, it follows that $\nn \hat{\bSigma}_n^\lambda-\bSigma_0 \nn_F = O_P(\sqrt{\frac{r (p^2 + q^2 + \log M)}{n}})=o_p(1)$. This asymptotic MSE convergence rate of the estimated covariance to the true covariance reflects the number of degrees of freedom of the model, which is on the order of the total number $r (p^2 + q^2)$ of unknown parameters. This result extends the recent high-dimensional results obtained in \cite{KGlasso13, TsiligkaridisTSP, TsiligkaridisSSP2012} for the single Kronecker product model (i.e., $r=1$).

Recall that $r \leq r_0 = \min(p^2,q^2)$. For the case when $p\sim q$, and $r\sim r_0$, we have a fully saturated Kronecker product model and the number of model parameters are of the order $p^4 \sim d^2$, and the SCM convergence rate (\ref{eq: SCM_rate}) coincides with the rate obtained in Thm. \ref{thm: PRLS_Frob_rate}.

For covariance models of low separation rank-i.e., $r \ll r_0$, Thm. \ref{thm: PRLS_Frob_rate} asserts that the high dimensional MSE convergence rate of PRLS can be much lower than the naive SCM convergence rate. Thus PRLS is an attractive alternative to rank-based series expansions like principal component analysis (PCA). We note that each term in the expansion $\bA_{0,\gamma} \otimes \bB_{0,\gamma}$ can be full-rank, while each term in the standard PCA expansion is rank 1.

Finally, we observe that Thm. \ref{thm: PRLS_Frob_rate} captures the tradeoff between estimation error and approximation error. In other words, choosing a smaller $r$ than the true separation rank would incur a larger approximation error $\inf_{\left\{\bR: \rank(\bR) \leq r\right\}} \nn \bR-\bR_0 \nn_F^2 > 0$, but smaller estimation error on the order of $O_P(\frac{r(p^2 +q^2 + \log M)}{n})$.


\subsection{Approximation Error}

It is well known from least-squares approximation theory that the residual error can be rewritten as:
\begin{equation} \label{eq:approx_error}
	\inf_{\bR:\rank(\bR)\leq r} \nn \bR-\bR_0 \nn_F^2 = \sum_{k=r+1}^{r_0} \sigma_k^2(\bR_0),
\end{equation}
where $\{\sigma_k(\bR_0)\}$ are the singular values of $\bR_0$. In the high dimensional setting, the sample size $n$ grows with the  dimensions $p,q$ so that the maximum separation rank $r_0$ also grows to infinity, and the approximation error (\ref{eq:approx_error}) may not be finite. In this case the bound in Theorem \ref{thm: PRLS_Frob_rate} will not be finite. Hence, an additional condition will be needed to ensure that the sum (\ref{eq:approx_error}) remains finite as $p,q\to\infty$: the singular values of $\bR_0$ need to decay faster than $O(1/k)$.

We show next that the class of block-Toeplitz covariance matrices have bounded approximation error if the separation rank scales like $\log(\max(p,q))$. To show this, we first provide a tight variational bound on the singular value spectrum of any $p^2 \times q^2$ matrix $\bR$. Note that the work on high dimensional Toeplitz covariance estimation under operator and Frobenius norms \cite{CaiZhangZhou:2010, BickelLevina:2008} are not applicable to the block-Toeplitz case. To establish Thm. \ref{thm: approx_err_blk_toep} on block Toeplitz matrices we first need the following Lemma.

\begin{lemma} (Variational Bound on Singular Value Spectrum) \label{lemma: variational_bnd}
	Let $\bR$ be an arbitrary matrix of size $p^2 \times q^2$. Let $\bP_k$ be an orthogonal projection of $\RR^{q^2}$ onto $\RR^k$. Then, for $k=1,\dots,r_0-1$ we have:
	\begin{equation} \label{eq: variational_bnd}
		\sigma_{k+1}^2(\bR) \leq \nn (\bI_{q^2}-\bP_k) \bR^T \nn_2^2
	\end{equation}
	with equality iff $\bP_k=\bV_k\bV_k^T$. Also, $\bV_k=[\bv_1,\dots,\bv_k]$, where $\bv_i$ is the $i$th column of $\bV$ and $\bR=\bU \bSigma \bV^T$ is the singular value decomposition. 
\end{lemma}
\begin{IEEEproof}
	See Appendix F.
\end{IEEEproof}

Using this fundamental lemma, we can characterize the approximation error for estimating block-Toeplitz matrices with exponentially decaying off-diagonal norms. Such matrices arise, for example, as covariance matrices of multivariate stationary random processes of dimension $m$ (see (\ref{eq:VAR})) and take the block Toeplitz form:
\begin{equation} \label{eq:block_toep}
	\underbrace{\bSigma_0}_{(N+1)m\times (N+1)m} = \begin{bmatrix} \bSigma(0) & \bSigma(1) & \dots & \bSigma(N) \\ \bSigma(-1) & \bSigma(0) & \dots & \bSigma(N-1) \\ \vdots & \vdots & \ddots & \vdots \\ \bSigma(-N) & \bSigma(-N+1) & \dots & \bSigma(0)  \end{bmatrix}
\end{equation}
where each submatrix is of size $m \times m$. For a zero-mean vector process $\by = \{\by(0), \dots, \by(N)\}$, the submatrices are given by $\bSigma(\tau) = \E[\by(0) \by(\tau)^T]$.

\begin{figure}[ht]
	\centering
		\includegraphics[width=0.80\textwidth]{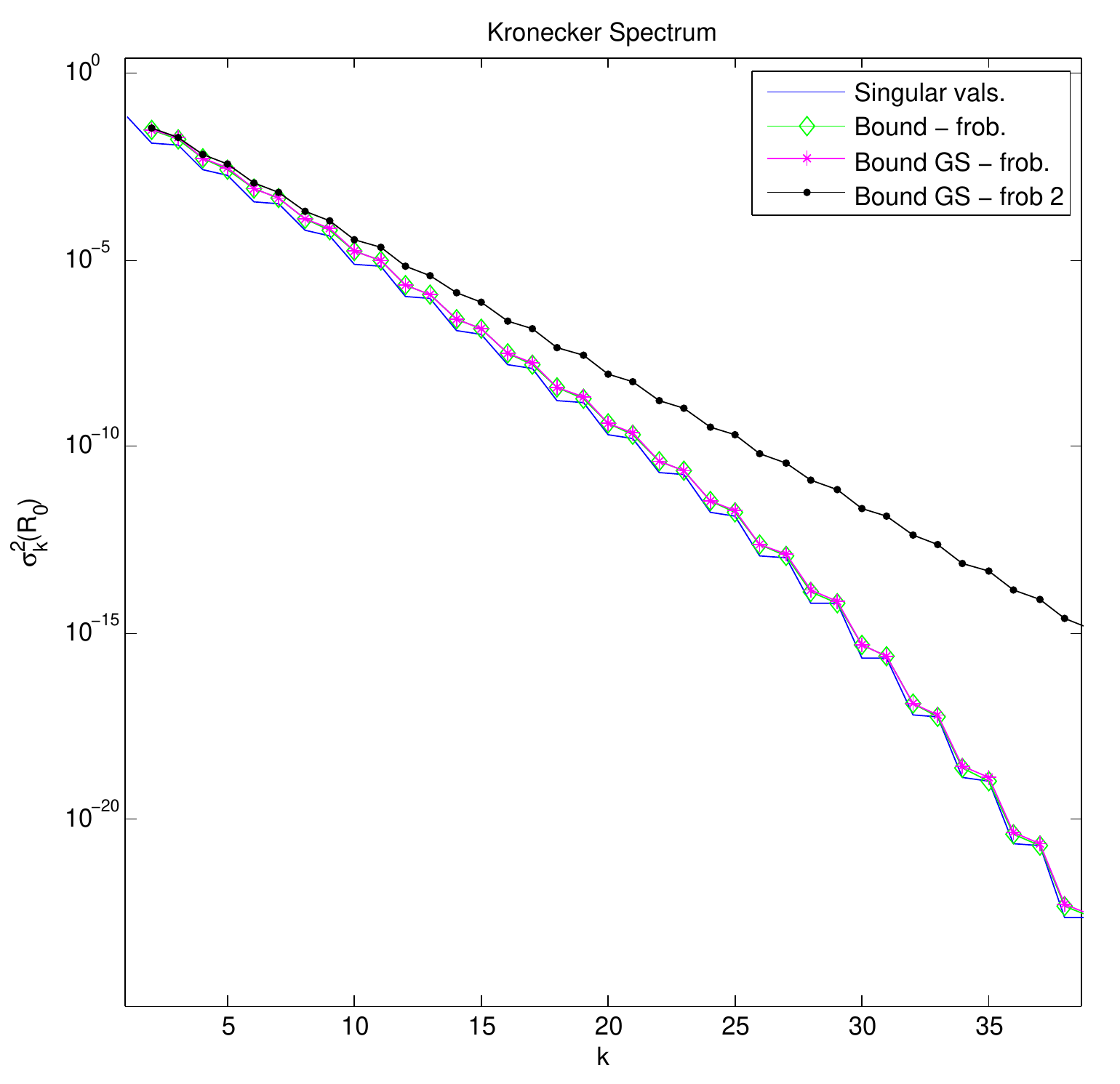}
	\caption{\label{fig:kron_spectrum_bnd} Kronecker spectrum and bounds based on Lemma \ref{lemma: variational_bnd}. The upper bound `Bound - frob' (in green) is obtained using the bound (\ref{eq: variational_bnd}) using the basis associated with the minimum $\ell_2$ approximation error (i.e., the optimal basis computed by SVD as outlined in the equality condition of Lemma \ref{lemma: variational_bnd}). The upper bound `Bound GS - frob' (in magenta) is constructed using the variational bound (\ref{eq: variational_bnd}) with projection matrix $\bP_k$ having columns drawn from the  orthonormal basis constructed in the proof of Thm. \ref{thm: approx_err_blk_toep}. The upper bound `Bound GS - frob 2' (in black) is constructed from the bound (\ref{eq:row_subtract}) in the proof of Thm. \ref{thm: approx_err_blk_toep}. }
\end{figure}

\begin{theorem} \label{thm: approx_err_blk_toep}
	Consider a block-Toeplitz p.d. matrix $\bSigma_0$ of size $(N+1)m \times (N+1)m$, with $\nn \bSigma(\tau) \nn_F^2 \leq C' u^{2|\tau|} q$ for all $\tau=-N,\dots,N$ and constant $u\in (0,1)$. Let $\hat{\bSigma}_n^\lambda$ be the de-permuted matrix $\mathcal{R}^{-1}(\hat{\bR}_n^\lambda)$, where $\hat{\bR}_n^\lambda$ is given in (\ref{eq:solution}). Using the minimal separation rank $r$:
	\begin{equation*}
		r \geq \frac{\log(pq/\epsilon)}{\log(1/u)}.
	\end{equation*}
	Then, the PRLS algorithm estimates $\bSigma_0$ up to an absolute tolerance $\epsilon \in (0,1)$ with convergence rate guarantee:
	\begin{equation}
		\nn \hat{\bSigma}_n^{\lambda} -\bSigma_0 \nn_F^2 \leq \epsilon + C' r \frac{p^2+q^2+\log M}{n}
	\end{equation}
	holding with probability at least $1-\max(p,q,n)^{-t/4C}$ for $\lambda$ chosen as perscribed in Thm. \ref{thm: PRLS_Frob_rate}. Here, $t>1$ is constant and $C,C'>0$ are constants specified in Thm. \ref{thm: PRLS_Frob_rate}.
\end{theorem}
\begin{IEEEproof}
	See Appendix G.
\end{IEEEproof}
The exponential norm decay condition of Thm. \ref{thm: approx_err_blk_toep} is satisfied by a first-order vector autoregressive process:
\begin{equation} \label{eq:VAR}
	\bZ_{t} = \Phi \bZ_{t-1}  + \Epsilon_t
\end{equation}
with $u = \nn \Phi \nn_2 \in (0,1)$, where $\bZ_t \in \RR^m$. For $\Epsilon_t \sim N(0,\bSigma_{\epsilon})$, this is a multivariate Gaussian process. Collecting data over a time horizon of size $N+1$, we concatenate these observations into a large random vector $\bz$ of dimension $(N+1)m$, where $m$ is the process dimension. The resulting covariance matrix has the block-Toeplitz form assumed in Thm. \ref{thm: approx_err_blk_toep}. Figure \ref{fig:kron_spectrum_bnd} shows bounds constructed using the Frobenius upper bound on the spectral norm in (\ref{eq: variational_bnd}) and using the projection matrix $\bP_k$ as discussed in the proof of Thm. \ref{thm: approx_err_blk_toep}. The bound given in the proof of Thm. \ref{thm: approx_err_blk_toep} (in black) is shown to be linear in log-scale, thus justifying the exponential decay of the Kronecker spectrum.

\section{Simulation Results} \label{sec: simulations}

We consider dense positive definite matrices $\bSigma_0$ of dimension $d=625$. Taking $p=q=25$, we note that the number of free parameters that describe each Kronecker product is of the order $p^2 + q^2 \sim p^2$, which is essentially of the same order as the number of unknown parameters required to specify each eigenvector of $\bSigma_0$, i.e., $pq \sim p^2$.

\subsection{Sum of Kronecker Product Covariance}
\begin{figure}[htp]
	\centering
		\includegraphics[width=0.80\textwidth]{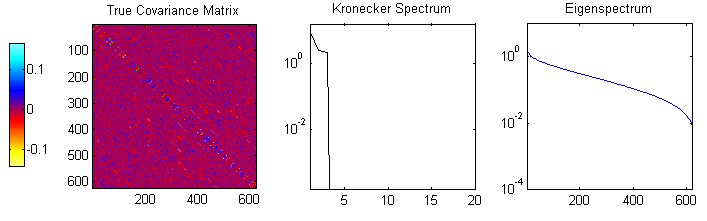}
	\caption{\label{fig:cov_kp} Simulation A. True dense covariance is constructed using the sum of KP model (\ref{eq: KP_model}), with $r=3$. Left panel: True positive definite covariance matrix $\bSigma_0$. Middle panel: Kronecker spectrum (eigenspectrum of $\bSigma_0$ in permuted domain). Right panel: Eigenspectrum (Eigenvalues of $\bSigma_0$). Note that the Kronecker spectrum is much more concentrated than the eigenspectrum. }
	\vfill
		\includegraphics[width=0.80\textwidth]{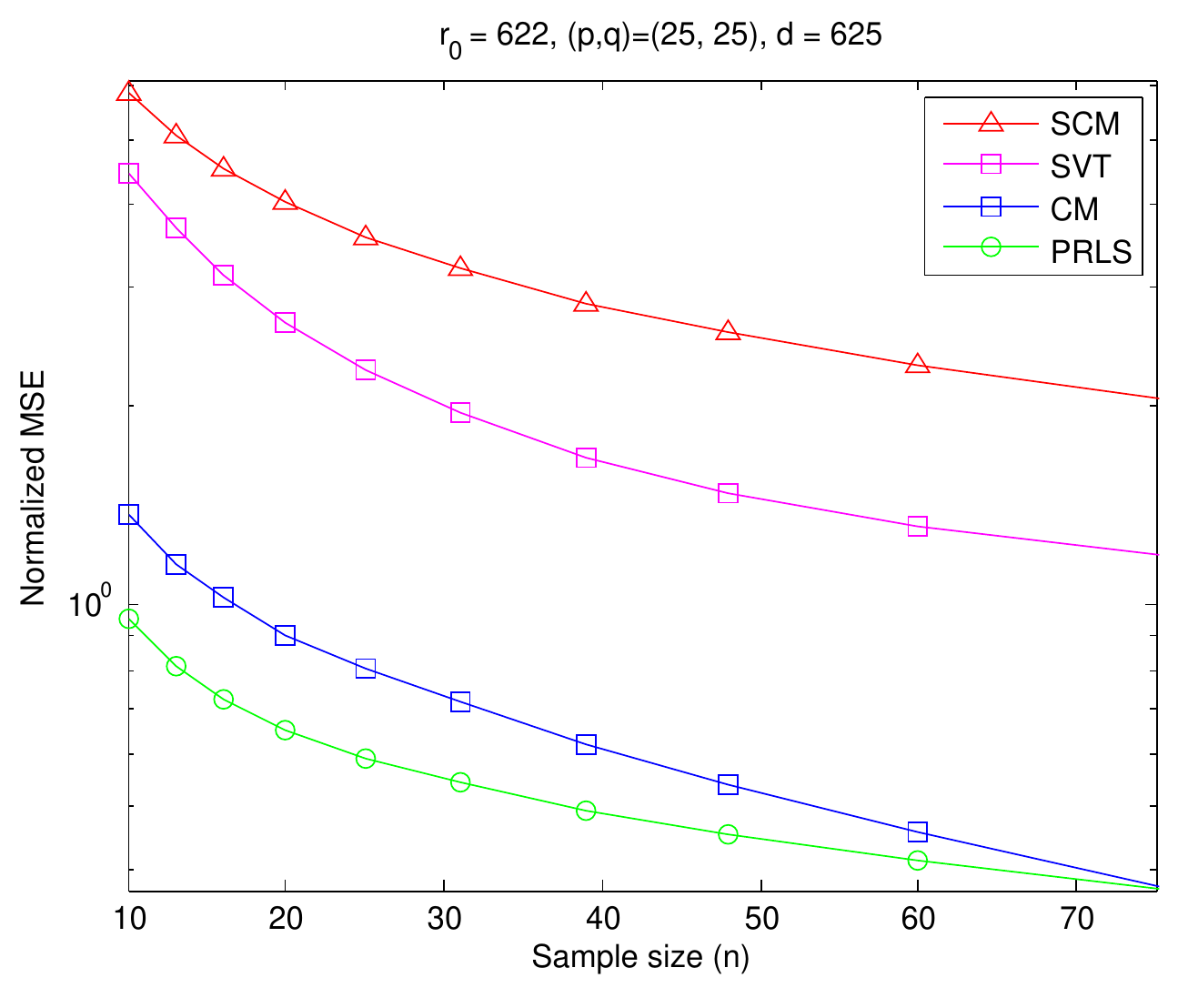}
	\caption{\label{fig:mse_kp} Simulation A. Normalized MSE performance for true covariance matrix in Fig.~\ref{fig:cov_kp} as a function of sample size $n$. PRLS outperforms CM, SVT (i.e., solution of (\ref{eq:SVT})) and the standard SCM estimator. Here, $p=q=25$ and $N_{MC}=80$. For $n=20$, PRLS achieves a $7.91$ dB MSE reduction over SCM and SVT achieves a $1.80$ dB MSE reduction over SCM. }
\end{figure}

The covariance matrix shown in Fig.~\ref{fig:cov_kp} was constructed using (\ref{eq: KP_model}) with $r=3$, with each p.d. factor chosen as $\bC\bC^T$, where $\bC$ is a square Gaussian random matrix. Fig.~\ref{fig:mse_kp} shows the empirical performance of covariance matching (CM) (i.e., solution of (\ref{eq: vanloan}) with $r=3$), PRLS and SVT (i.e., solution of (\ref{eq:SVT})). We note that the Kronecker spectrum contains only three nonzero terms while the true covariance is full rank. The PRLS spectrum is more concentrated than the eigenspectrum and, from Fig. \ref{fig:mse_kp}, we observe PRLS outperforms covariance matching (CM), SVT and SCM across all $n$.

\subsection{Block Toeplitz Covariance}
\begin{figure}[htp]
	\centering
		\includegraphics[width=0.80\textwidth]{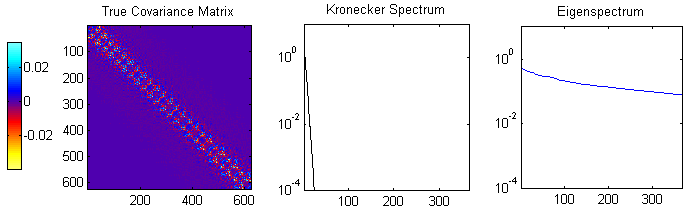}
	\caption{\label{fig:cov} Simulation B. True dense block-Toeplitz covariance matrix. Left panel: True positive definite covariance matrix $\bSigma_0$. Middle panel: Kronecker spectrum (eigenspectrum of $\bSigma_0$ in permuted domain). Right panel: Eigenspectrum (Eigenvalues of $\bSigma_0$). Note that the Kronecker spectrum is much more concentrated than the eigenspectrum. }
	\vfill
		\includegraphics[width=0.80\textwidth]{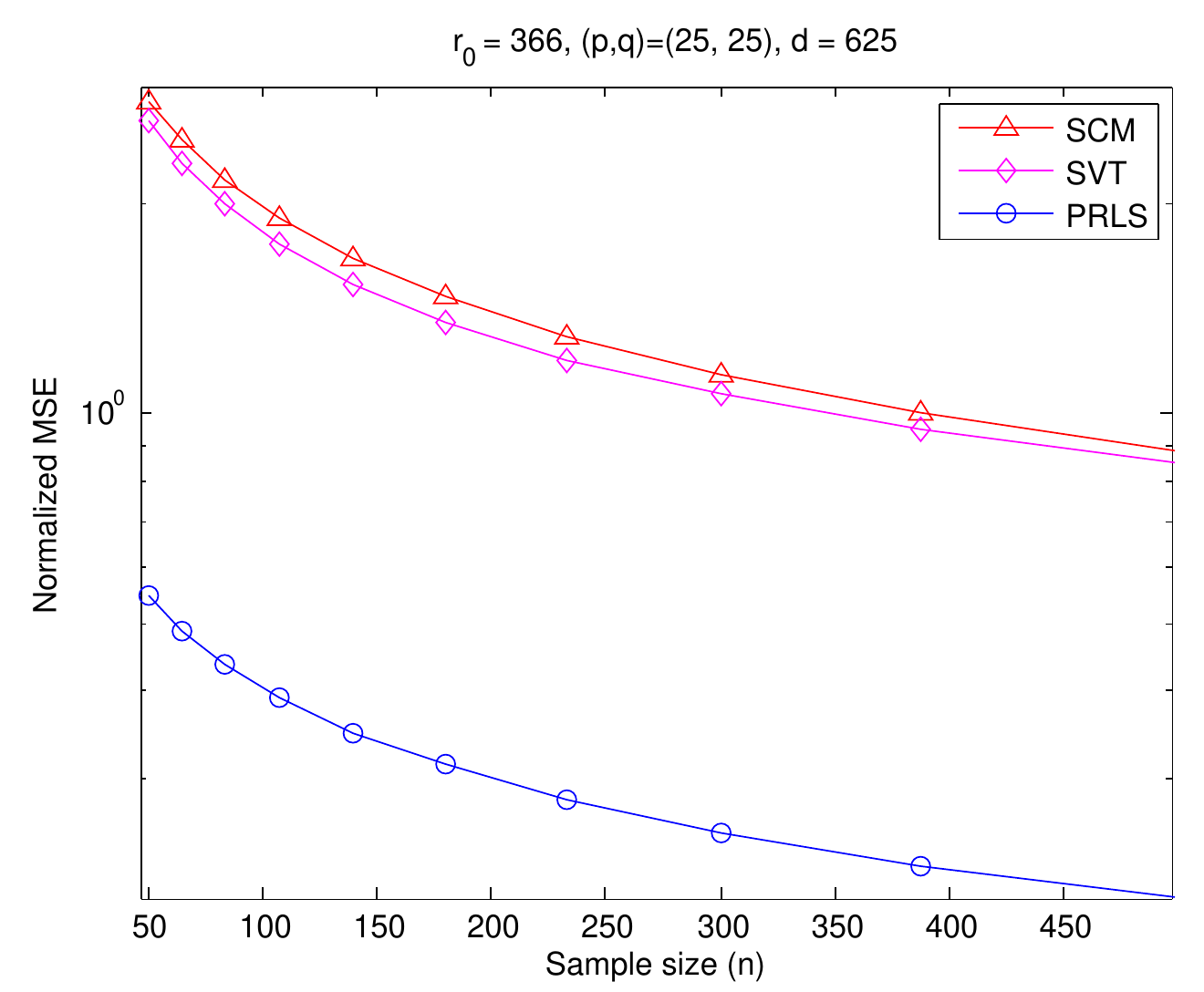}
	\caption{\label{fig:mse_blk_toep} Simulation B. Normalized MSE performance for covariance matrix in Fig.~\ref{fig:cov} as a function of sample size $n$. PRLS outperforms SVT (i.e., solution of (\ref{eq:SVT})) and the standard SCM estimator. Here, $p=q=25$ and $N_{MC}=80$. For $n=108$, PRLS achieves a $6.88$ dB MSE reduction over SCM and SVT achieves a $0.37$ dB MSE reduction over SCM. Note again that the Kronecker spectrum is much more concentrated than the eigenspectrum. }
\end{figure}

The covariance matrix shown in Fig.~\ref{fig:cov} was constructed by first generating a Gaussian random square matrix $\Phi$ of spectral norm $0.95<1$, and then simulating the block Toeplitz covariance for the process shown in (\ref{eq:VAR}). Fig.~\ref{fig:mse_blk_toep} compares the empirical performance of PRLS and SVT  (i.e., the solution of (\ref{eq:SVT}) with appropriate scaling for the regularization parameter). We observe that the Kronecker product estimator performs much better than both SVT (i.e., the solution of (\ref{eq:SVT})) and naive SCM estimator. This is most likely due to the fact that the repetitive block structure of Kronecker products better summarizes the covariance structure. We observe from Fig. \ref{fig:cov} that for this block Toeplitz covariance, the Kronecker spectrum decays more rapidly (exponentially) than the eigenspectrum.

\section{Application to Wind Speed Prediction} \label{sec:real_data}

In this section, we demonstrate the performance of PRLS in a real world application: wind speed prediction. We apply our methods to the Irish wind speed dataset and the NCEP dataset.

\subsection{Irish Wind Speed Data}
We use data consisting of time series consisting of daily average wind speed recordings during the period $1961 - 1978$ at $q=11$ meteorological stations. This data set has many temporal coordinates, spanning a total of $n_{total}=365 \cdot 8=2920$ daily average recordings of wind speed at each station. More details on this data set can be found in \cite{Haslett:1989, Gneiting:2002, deLunaGenton:2005, Stein:2005} and it can be downloaded from Statlib \text{http://lib.stat.cmu.edu/datasets}. We used the same square root transformation, estimated seasonal effect offset and station-specific mean offset as in \cite{Haslett:1989}, yielding the multiple (11) velocity measures. We used the data from years $1969-1970$ for training and the data from $1971-1978$ for testing.

The task is to predict the average velocity for the next day using the average wind velocity in each of the $p-1$ previous days. The full dimension of each observation vector is $d=pq$, and each $d$-dimensional observation vector is formed by concatenating the $p$ time-consecutive $q$-dimensional vectors (each entry containing the velocity measure for each station) without overlapping the time segments. The SCM was estimated using data from the training period consisting of years $1969-1970$. Linear predictors over the time series were constructing by using these estimated covariance matrices in an ordinary least squares predictor. Specifically, we constructed the SCM linear predictor of all stations' wind velocity from the $p-1$ previous samples of the $q=11$ stations' time series:
\begin{equation} \label{eq:linear_predictor}
	\hat{\bv}_t = \bSigma_{2,1} \bSigma_{1,1}^{-1} \bv_{t-1:t-(p-1)}
\end{equation}
where $\bv_{t-1:t-(p-1)} \in \RR^{(p-1)q}$ is the stacked wind velocities from the previous $p-1$ time instants and $\bSigma_{2,1} \in \RR^{q\times q(p-1)}$ and $\bSigma_{1,1}\in \RR^{q(p-1)\times q(p-1)}$ are submatrices of the $qp \times qp $ standard SCM:
\begin{equation*}
	\hat{\bS}_n = \begin{bmatrix} \bSigma_{1,1} & \bSigma_{1,2} \\ \bSigma_{2,1} & \bSigma_{2,2} \end{bmatrix}
\end{equation*}
The PRLS predictor was similarly constructed using our proposed estimator of the $qp \times qp$ Kronecker sum covariance matrix instead of the SCM. The coefficients of each of these predictors, $\bSigma_{2,1} \bSigma_{1,1}^{-1}$, were subsequently applied to predict over the test set. 

The predictors were tested on the data from years $1971-1978$, corresponding to $n_{test}=365\cdot8 = 2920$ days, as the ground truth. Using non-overlapping samples and $p=8$, we have a total of $n = \lceil \frac{365 \cdot 2}{p} \rceil = 91$ training samples of full dimension $d=88$.

\begin{figure}[ht]
	\centering
		\includegraphics[width=0.70\textwidth]{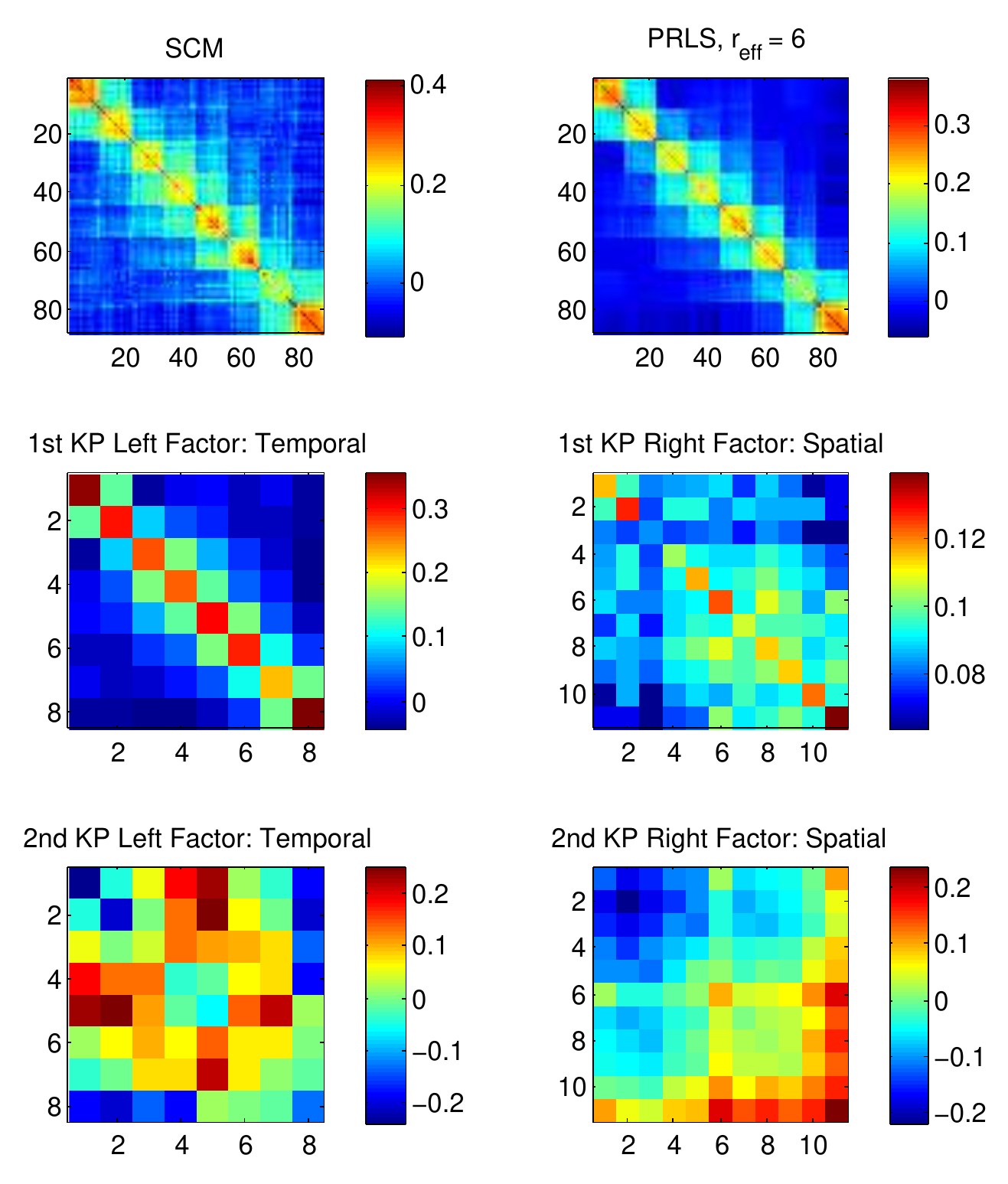}
	\caption{\label{fig:Irish_approx} Irish wind speed data: Sample covariance matrix (SCM) (top left), PRLS covariance estimate (top right), temporal Kronecker factor for first KP component (middle left) and spatial Kronecker factor for first KP component (middle right), temporal Kronecker factor for second KP component (bottom left) and spatial Kronecker factor for second KP component (bottom right). Note that the second order factors are not necessarily positive definite, although the sum of the components (i.e., the PRLS solution) is positive definite for large enough $n$. Each KP factor has unit Frobenius norm. Note that the plotting scales the image data to the full range of the current colormap to increase visual contrast.  }
\end{figure}

Fig.~\ref{fig:Irish_approx} shows the Kronecker product factors that make up the solution of Eq. (\ref{eq: vanloan}) with $r=1$ and the PRLS estimate. The PRLS estimate contains $r_{eff}=6$ nonzero terms in the KP expansion. It is observed that the first order temporal factor gives a decay in correlations over time, and spatial correlations between weather stations are present. The second order temporal and spatial factors can potentially give insight into long range dependencies.
\begin{figure}[ht]
	\centering
		\includegraphics[width=1.00\textwidth]{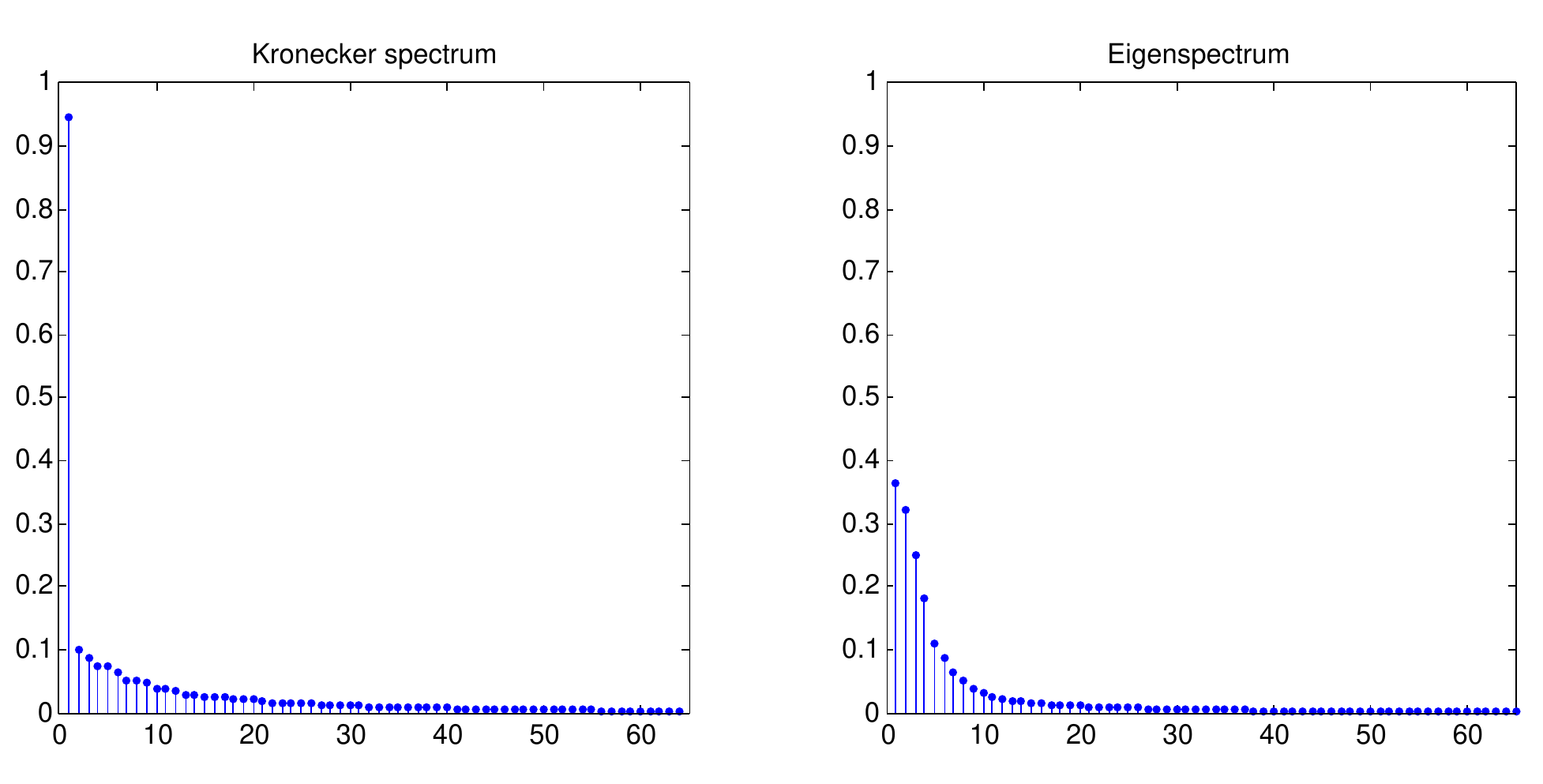}
	\caption{\label{fig:Irish_spectra} Irish wind speed data: Kronecker spectrum of SCM (left) and Eigenspectrum of SCM (right). The first and second KP components contain $94.60 \%$ and $1.07 \%$ of the spectrum energy. The first and second eigenvectors contain $36.28 \%$ and $28.76 \%$ of the spectrum energy. The KP spectrum is more compact than the eigenspectrum. Here, the eigenspectrum is truncated at $\min(p^2,q^2)=8^2=64$ to match the Kronecker spectrum. Each spectrum was normalized such that each component has height equal to the percentage of energy associated with it. }
\end{figure}

\begin{figure}[ht]
	\centering
		\includegraphics[width=1.00\textwidth]{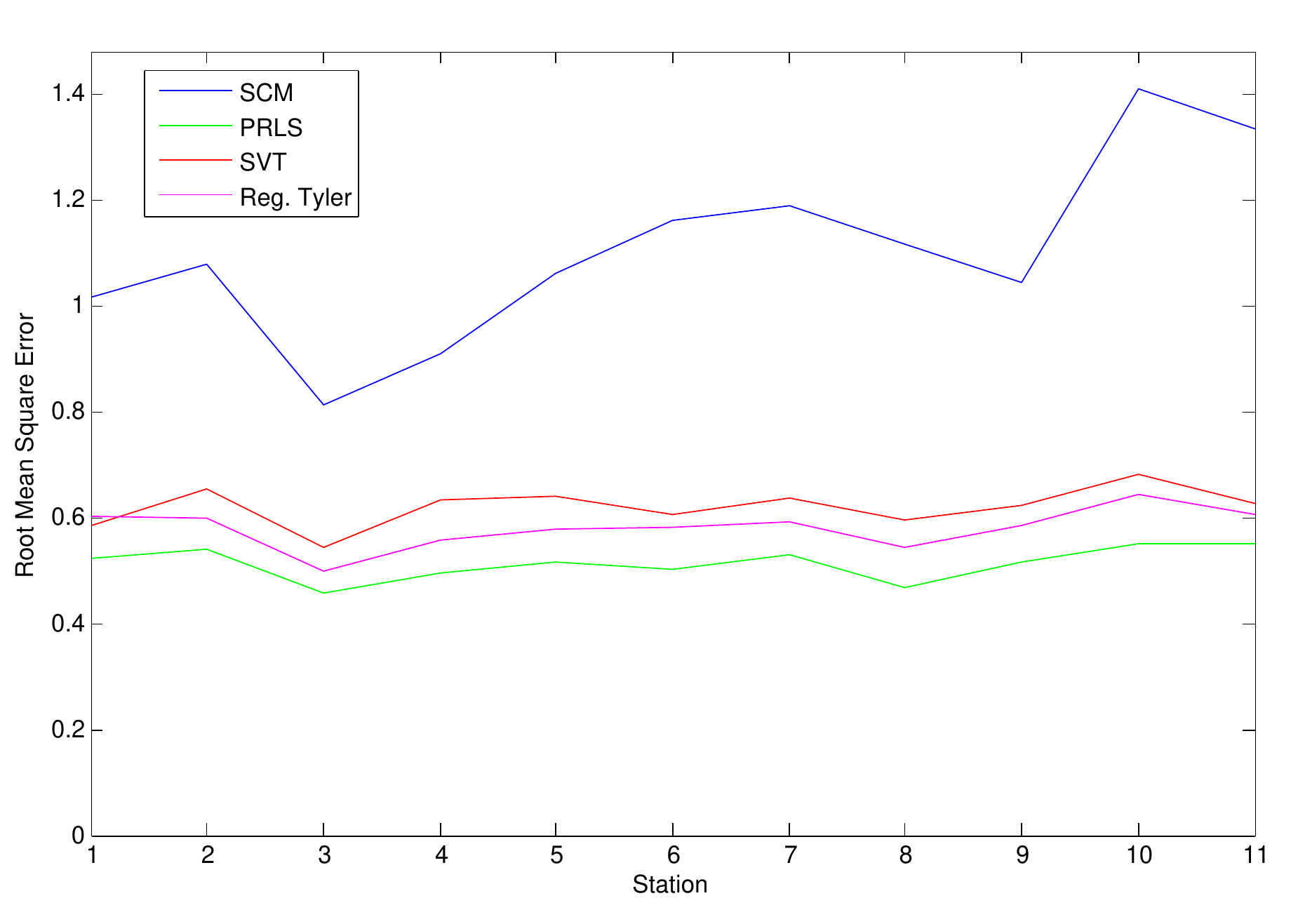}
	\caption{\label{fig:Irish_rmse_pred} Irish wind speed data: RMSE prediction performance across $q$ stations for linear estimators using SCM (blue), PRLS (green), SVT (red) and regularized Tyler (magenta). PRLS, SVT and regularized Tyler respectively achieve an average reduction in RMSE of $3.32$, $2.50$ and $2.79$ dB as compared to SCM (averaged across stations). }
\end{figure}

\begin{figure}[ht]
	\centering
		\includegraphics[width=1.00\textwidth]{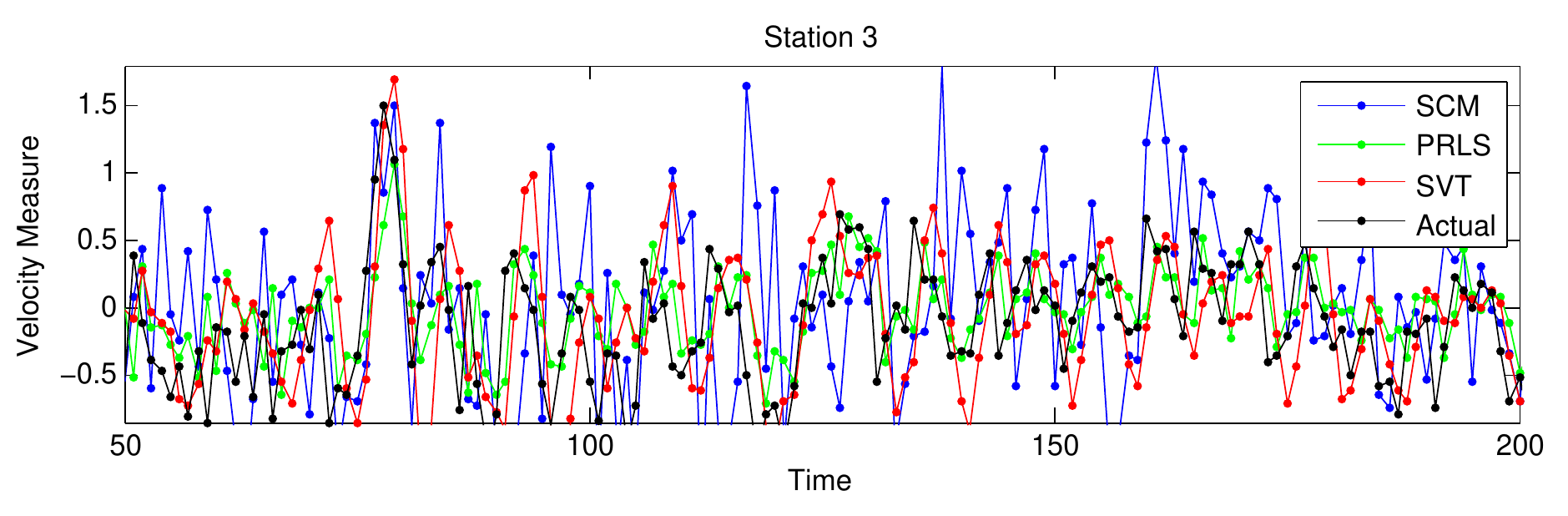}
	\caption{\label{fig:Irish_pred} Irish wind speed data: Prediction performance for linear estimators using SCM (blue), SVT (red) and PRLS (green) for a time interval of $150$ days. The actual (ground truth) wind speeds are shown in black. PRLS offers better tracking performance as compared to SVT and SCM. }
\end{figure}

Fig.~\ref{fig:Irish_rmse_pred} shows the root mean squared error (RMSE) prediction performance over the testing period of $2920$ days for the forecasts based on the standard SCM, PRLS estimator, Lounici's SVT estimator \cite{Lounici}, and regularized Tyler \cite{ChenWieselHero:2011}. The PRLS estimator was implemented using a regularization parameter $\lambda_n = C \nn \hat{\bS}_n \nn_2 \sqrt{\frac{p^2+q^2+\log(\max(p,q,n))}{n}}$ with $C=0.13$. The constant $C$ was chosen by optimizing the prediction RMSE on the training set over a range of regularization parameters $\lambda$ parameterized by $C$. The SVT estimator proposed by Lounici \cite{Lounici} was implemented using a regularization parameter $\lambda = C \sqrt{\tr(\hat{S}_n) \nn\hat{S}_n\nn_2} \sqrt{\frac{\log(2pq)}{n}}$ with constant $C=1.9$ optimized in a similar manner. The regularized Tyler estimator was implemented using the data-dependent shrinkage coefficient suggested in Eqn. (13) in \cite{ChenWieselHero:2011}. Fig.~\ref{fig:Irish_pred} shows a sample period of $150$ days. We observe that PRLS tracks the actual wind speed better than the SCM-based predictor does.


\subsection{NCEP Wind Speed Data}
We use data representative of the wind conditions in the lower troposphere (surface data at .995 sigma level) for the global grid ($90^\circ$N - $90^\circ$S, $0^\circ$E - $357.5^\circ$E). We obtained the data from the National Centers for Environmental Prediction reanalysis project (Kalnay et al. \cite{Kalnay:1996}), which is available online at the NOAA website \text{ftp://ftp.cdc.noaa.gov/Datasets/ncep.reanalysis.dailyavgs/surface}. Daily averages of U (east-west) and V (north-south) wind components were collected using a station grid of size $144 \times 73$ (2.5 degree latitude $\times$ 2.5 degree longitude global grid) over the years $1948-2012$. The wind speed is computed by taking the magnitude of the wind vector.

\subsubsection{Continental US Region}
\begin{figure}[ht]
	\centering
		\includegraphics[width=0.70\textwidth]{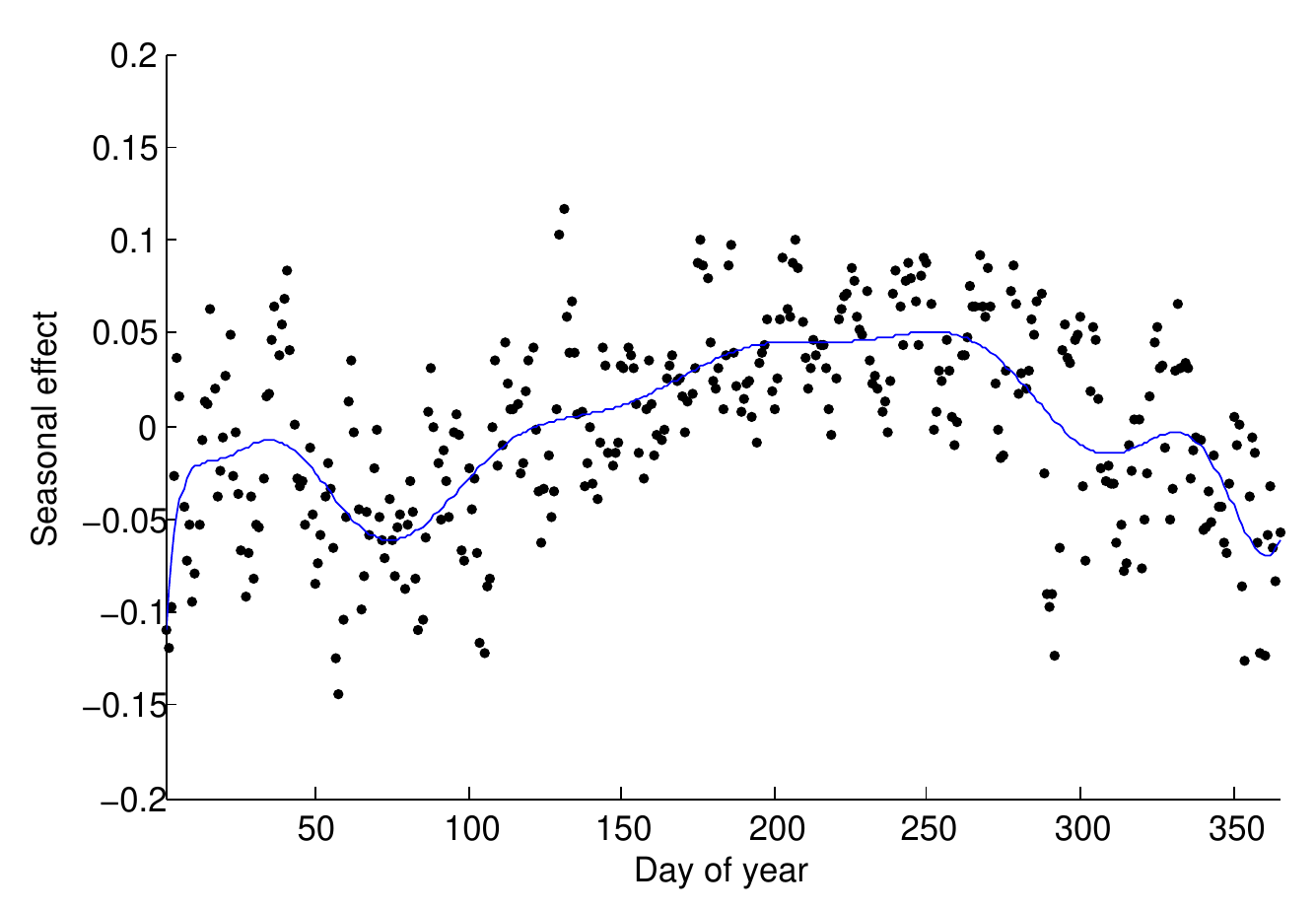}
	\caption{\label{fig:US_seasonal_effect} NCEP wind speed data (Continental US): Seasonal effect as a function of day of the year. A $14$th order polynomial is fit by the least squares method to the average of the square root of the daily mean wind speeds over all stations and over all training years. }
\end{figure}

We considered a $10\times 10$ grid of stations, corresponding to latitude range $25^\circ$N-$47.5^\circ$N and longitude range $125^\circ$W-$97.5^\circ$W. For this selection of variables, $q = 10 \cdot 10 = 100$ is the total number of stations and $p-1=7$ is the prediction time lag. We preprocessed the raw data using the detrending procedure outlined in Haslett et al. \cite{Haslett:1989}. More specifically, we first performed a square root transformation, then estimated and subtracted the station-specific means from the data and finally estimated and subtracted the seasonal effect (see Fig.~\ref{fig:US_seasonal_effect}). The resulting features/observations are called the velocity measures \cite{Haslett:1989}.
The SCM was estimated using data from the training period consisting of years $2003-2007$. Since the SCM is not full rank, the linear preictor (\ref{eq:linear_predictor}) was implemented with the Moore-Penrose pseudo-inverse of $\bSigma_{1,1}$. The predictors were tested on the data from years $2008-2012$ as the ground truth. Using non-overlapping samples and $p=8$, we have a total of $n = \lceil \frac{365 \cdot 5}{p} \rceil = 228$ training samples of full dimension $d=800$.

\begin{figure}[ht]
	\centering
		\includegraphics[width=0.70\textwidth]{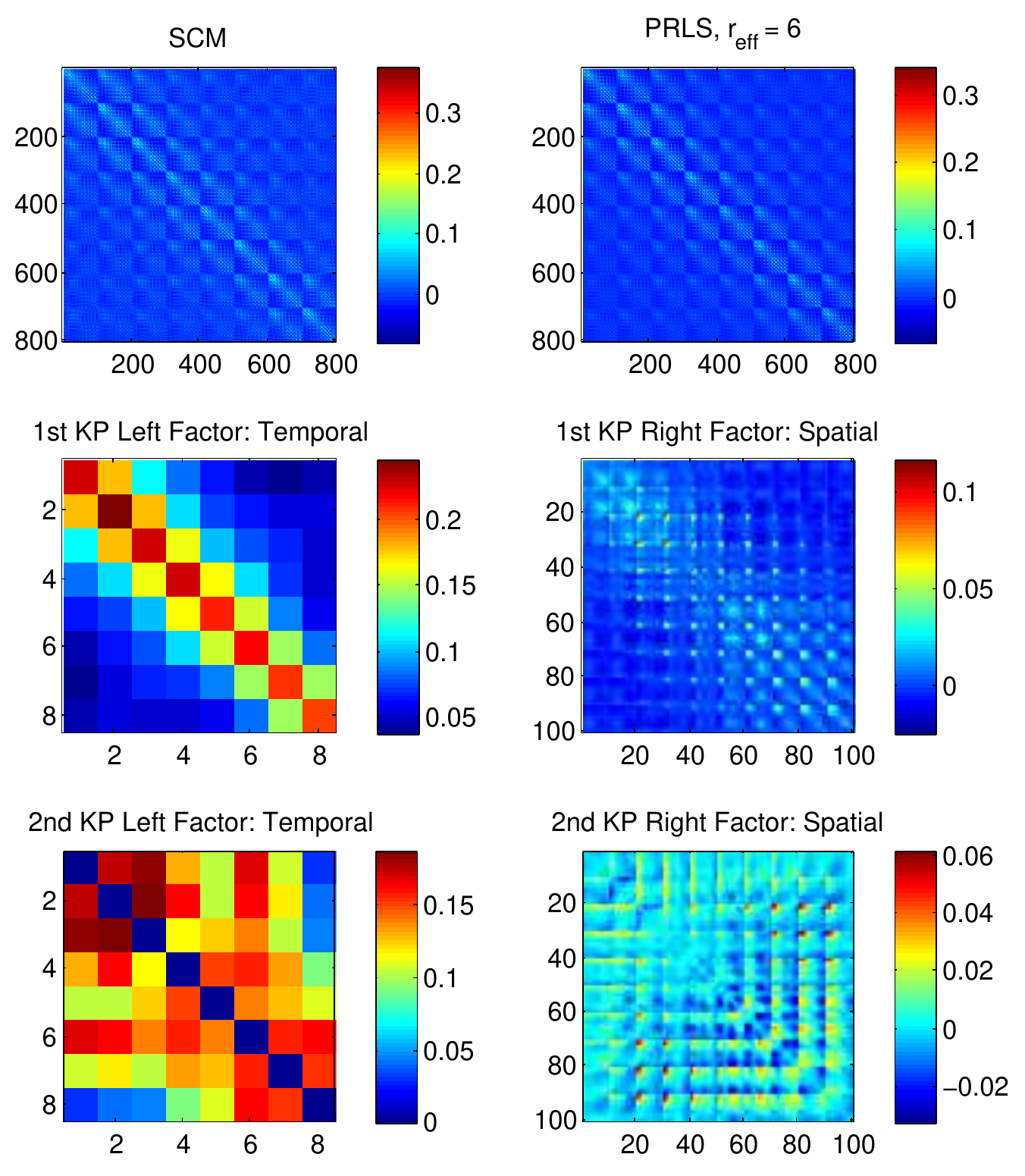}
	\caption{\label{fig:US_approx} NCEP wind speed data (Continental US): Sample covariance matrix (SCM) (top left), PRLS covariance estimate (top right), temporal Kronecker factor for first KP component (middle left) and spatial Kronecker factor for first KP component (middle right), temporal Kronecker factor for second KP component (bottom left) and spatial Kronecker factor for second KP component (bottom right). Note that the second order factors are not necessarily positive definite, although the sum of the components (i.e., the PRLS solution) is positive definite for large enough $n$. Each KP factor has unit Frobenius norm. Note that the plotting scales the image data to the full range of the current colormap to increase visual contrast.   }
\end{figure}

Fig.~\ref{fig:US_approx} shows the Kronecker product factors that make up the solution of Eq. (\ref{eq: vanloan}) with $r=2$ and the PRLS covariance estimate. The PRLS estimate contains $r_{eff}=6$ nonzero terms in the KP expansion. It is observed that the first order temporal factor gives a decay in correlations over time, and spatial correlations between weather stations are present. The second order temporal and spatial factors give some insight into longer range dependencies.
\begin{figure}[ht]
	\centering
		\includegraphics[width=1.00\textwidth]{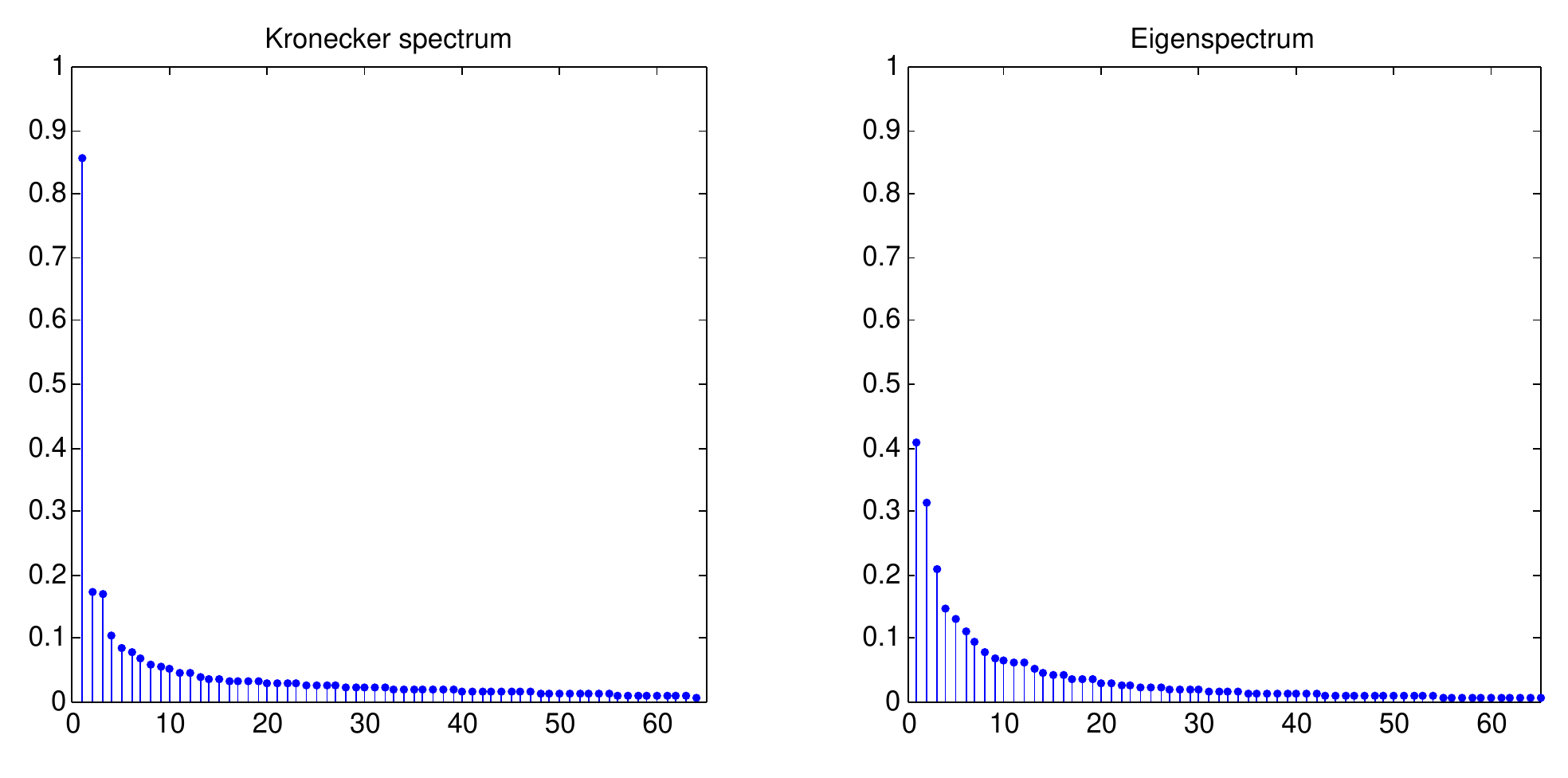}
	\caption{\label{fig:US_spectra} NCEP wind speed data (Continental US): Kronecker spectrum of SCM (left) and Eigenspectrum of SCM (right). The first and second KP components contain $85.88 \%$ and $3.48 \%$ of the spectrum energy. The first and second eigenvectors contain $40.93 \%$ and $23.82 \%$ of the spectrum energy. The KP spectrum is more compact than the eigenspectrum. Here, the eigenspectrum is truncated at $\min(p^2,q^2)=8^2=64$ to match the Kronecker spectrum. Each spectrum was normalized such that each component has height equal to the percentage of energy associated with it.  }
\end{figure}

\begin{figure}[ht]
	\centering
		\includegraphics[width=1.00\textwidth]{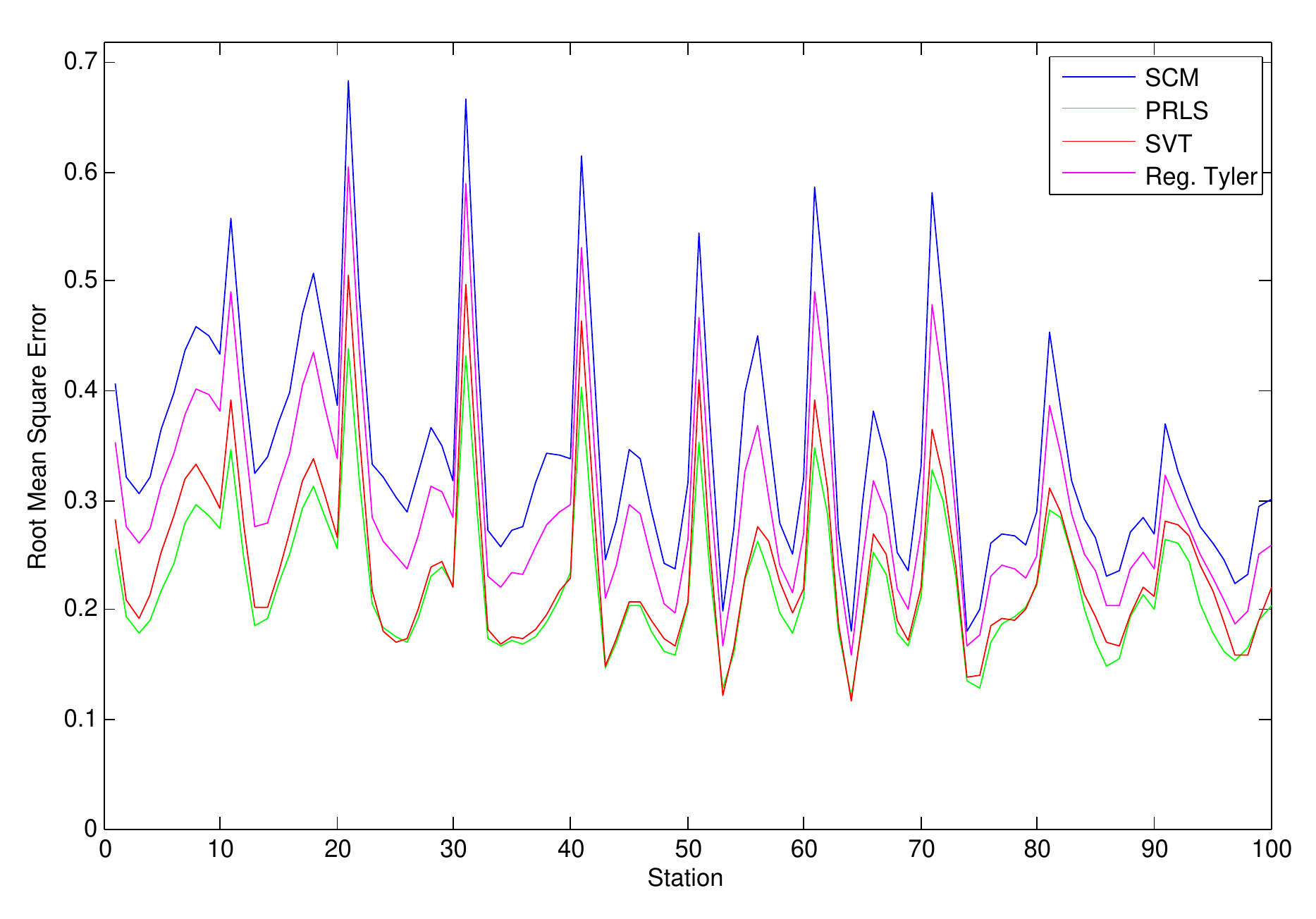}
	\caption{\label{fig:US_rmse_pred} NCEP wind speed data (Continental US): RMSE prediction performance across $q$ stations for linear estimators using SCM (blue), SVT (red), PRLS (green) and regularized Tyler (magenta). The estimators PRLS, SVT, and regularized Tyler  respectively achieve an average reduction in RMSE of $1.90$, $1.59$, and $0.66$ dB as compared to SCM (averaged across stations). }
\end{figure}

\begin{figure}[ht]
	\centering
		\includegraphics[width=1.00\textwidth]{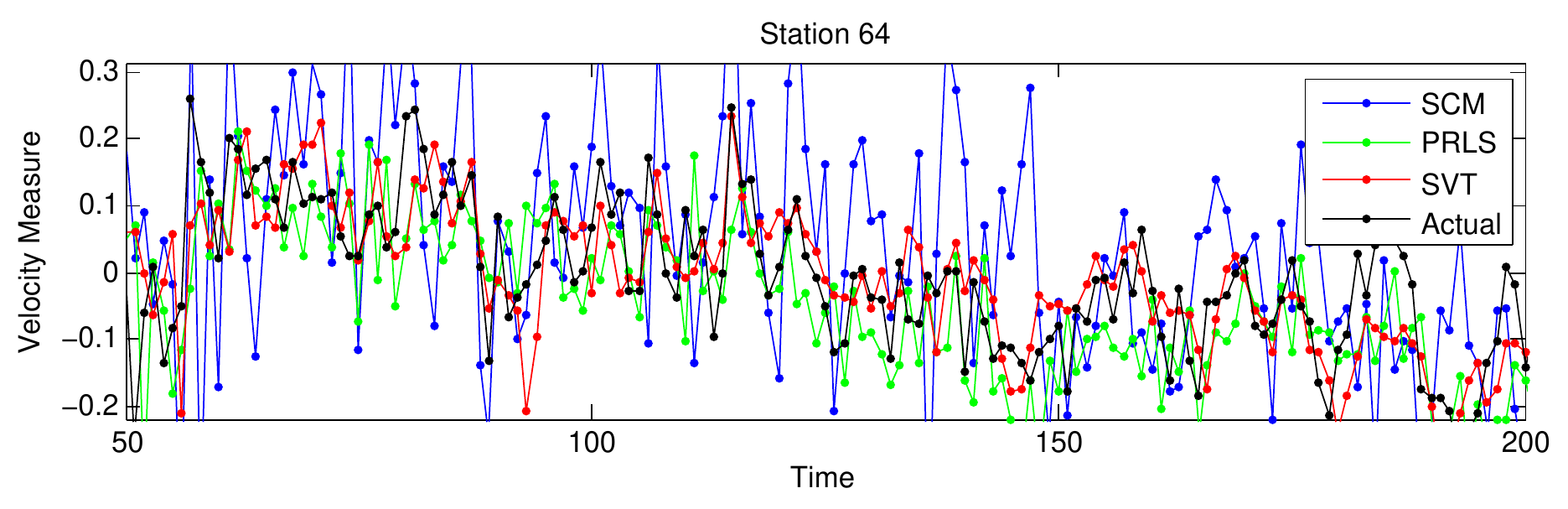}
	\caption{\label{fig:US_pred} NCEP wind speed data (Continental US): Prediction performance for linear estimators using SCM (blue), SVT (red) and PRLS (green) for a time interval of $150$ days. The actual (ground truth) wind speeds are shown in black. PRLS offers better tracking performance as compared to SCM and SVT. }
\end{figure}

Fig.~\ref{fig:US_rmse_pred} shows the root mean squared error (RMSE) prediction performance over the testing period of $1825$ days for the forecasts based on the standard SCM, PRLS, SVT \cite{Lounici} and regularized Tyler \cite{ChenWieselHero:2011}. The PRLS estimator was implemented using a regularization parameter $\lambda_n = C \nn \hat{\bS}_n \nn_2 \sqrt{\frac{p^2+q^2+\log(\max(p,q,n))}{n}}$ with $C=0.036$. The constant $C$ was chosen by optimizing the prediction RMSE on the training set over a range of regularization parameters $\lambda$ parameterized by $C$ (as in Irish wind speed data set). The SVT estimator proposed by Lounici \cite{Lounici} was implemented using a regularization parameter $\lambda = C \sqrt{\tr(\hat{S}_n) \nn\hat{S}_n\nn_2} \sqrt{\frac{\log(2pq)}{n}}$ with constant $C=0.31$ optimized in a similar manner. Fig.~\ref{fig:US_pred} shows a sample period of $150$ days. It is observed that SCM has unstable performance, while the Kronecker product estimator offers better tracking of the wind speeds.

\subsubsection{Arctic Ocean Region}
\begin{figure}[ht]
	\centering
		\includegraphics[width=0.70\textwidth]{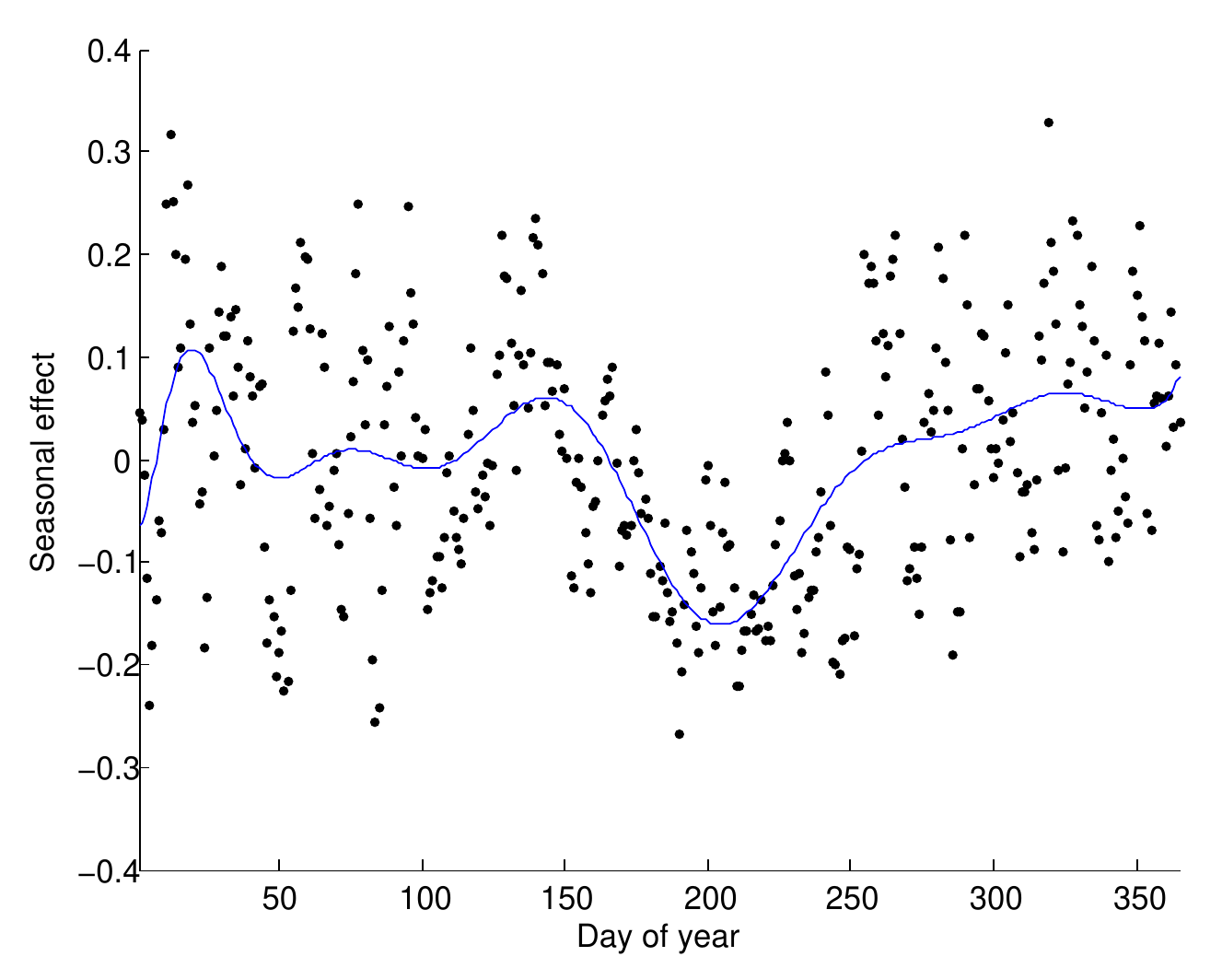}
	\caption{\label{fig:arctic_seasonal_effect} NCEP wind speed data (Arctic Ocean): Seasonal effect as a function of day of the year. A $14$th order polynomial is fit by the least squares method to the average of the square root of the daily mean wind speeds over all stations and over all training years. }
\end{figure}

We considered a $10\times 10$ grid of stations, corresponding to latitude range $90^\circ$N-$67.5^\circ$N and longitude range $0^\circ$E-$22.5^\circ$E. For this selection of variables, $q = 10 \cdot 10 = 100$ is the total number of stations and $p-1=7$ is the prediction time lag. We preprocessed the raw data using the detrending procedure outlined in Haslett et al. \cite{Haslett:1989}. More specifically, we first performed a square root transformation, then estimated and subtracted the station-specific means from the data and finally estimated and subtracted the seasonal effect (see Fig.~\ref{fig:arctic_seasonal_effect}). The resulting features/observations are called the velocity measures \cite{Haslett:1989}.
The SCM was estimated using data from the training period consisting of years $2003-2007$. Since the SCM is not full rank, the linear preictor (\ref{eq:linear_predictor}) was implemented with the Moore-Penrose pseudo-inverse of $\bSigma_{1,1}$. The predictors were tested on the data from years $2008-2012$ as the ground truth. Using non-overlapping samples and $p=8$, we have a total of $n = \lceil \frac{365 \cdot 5}{p} \rceil = 228$ training samples of full dimension $d=800$.

\begin{figure}[ht]
	\centering
		\includegraphics[width=0.70\textwidth]{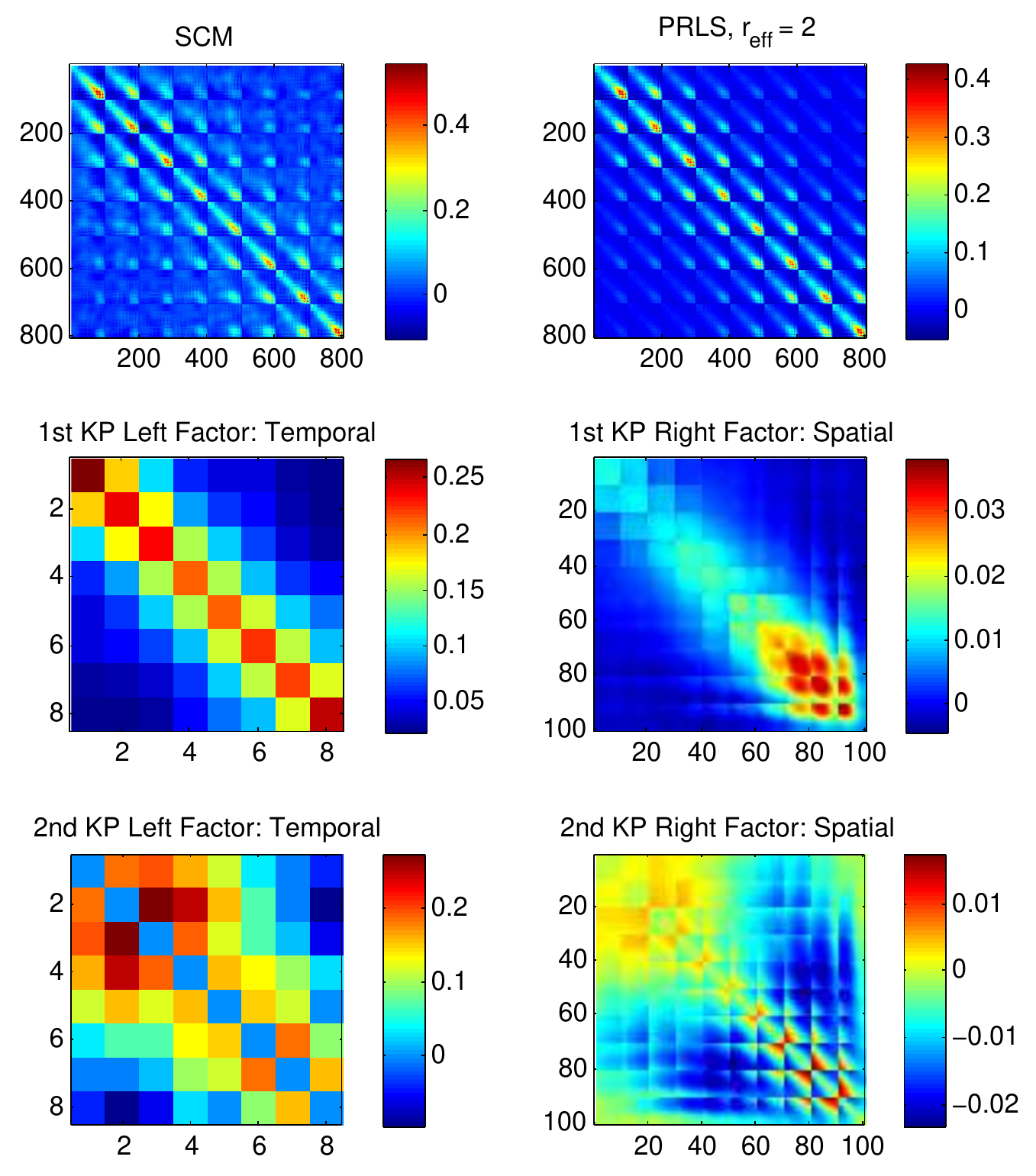}
	\caption{\label{fig:arctic_approx} NCEP wind speed data (Arctic Ocean): Sample covariance matrix (SCM) (top left), PRLS covariance estimate (top right), temporal Kronecker factor for first KP component (middle left) and spatial Kronecker factor for first KP component (middle right), temporal Kronecker factor for second KP component (bottom left) and spatial Kronecker factor for second KP component (bottom right). Note that the second order factors are not necessarily positive definite, although the sum of the components (i.e., the PRLS solution) is positive definite for large enough $n$. Each KP factor has unit Frobenius norm. Note that the plotting scales the image data to the full range of the current colormap to increase visual contrast.   }
\end{figure}

Fig.~\ref{fig:arctic_approx} shows the Kronecker product factors that make up the solution of Eq. (\ref{eq: vanloan}) with $r=2$ and the PRLS covariance estimate. The PRLS estimate contains $r_{eff}=2$ nonzero terms in the KP expansion. It is observed that the first order temporal factor gives a decay in correlations over time, and spatial correlations between weather stations are present. The second order temporal and spatial factors give some insight into longer range dependencies.
\begin{figure}[ht]
	\centering
		\includegraphics[width=1.00\textwidth]{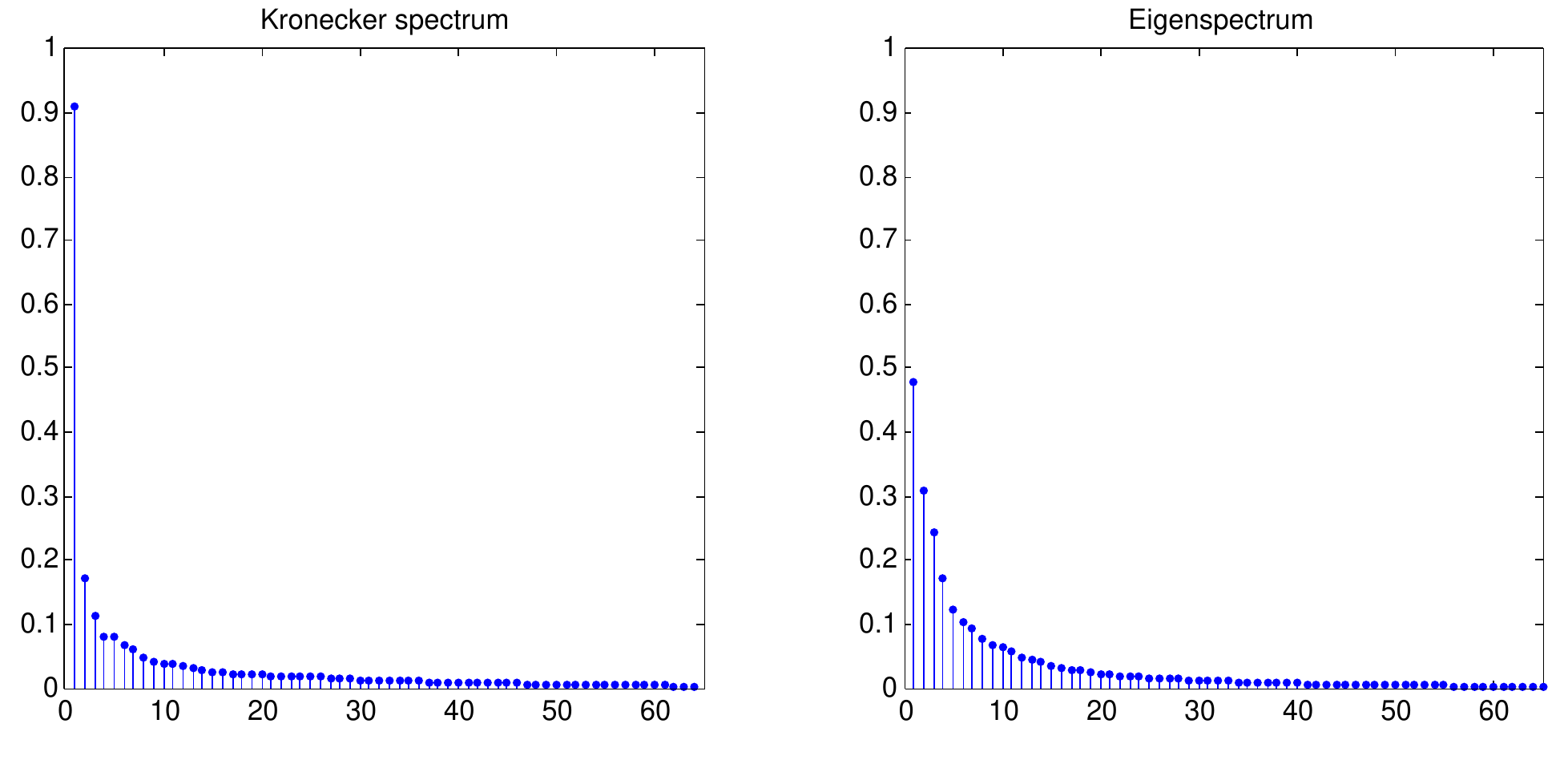}
	\caption{\label{fig:arctic_spectra} NCEP wind speed data (Arctic Ocean): Kronecker spectrum of SCM (left) and Eigenspectrum of SCM (right). The first and second KP components contain $91.12 \%$ and $3.28 \%$ of the spectrum energy. The first and second eigenvectors contain $47.99 \%$ and $19.68 \%$ of the spectrum energy. The KP spectrum is more compact than the eigenspectrum. Here, the eigenspectrum is truncated at $\min(p^2,q^2)=8^2=64$ to match the Kronecker spectrum. Each spectrum was normalized such that each component has height equal to the percentage of energy associated with it.  }
\end{figure}

\begin{figure}[ht]
	\centering
		\includegraphics[width=1.00\textwidth]{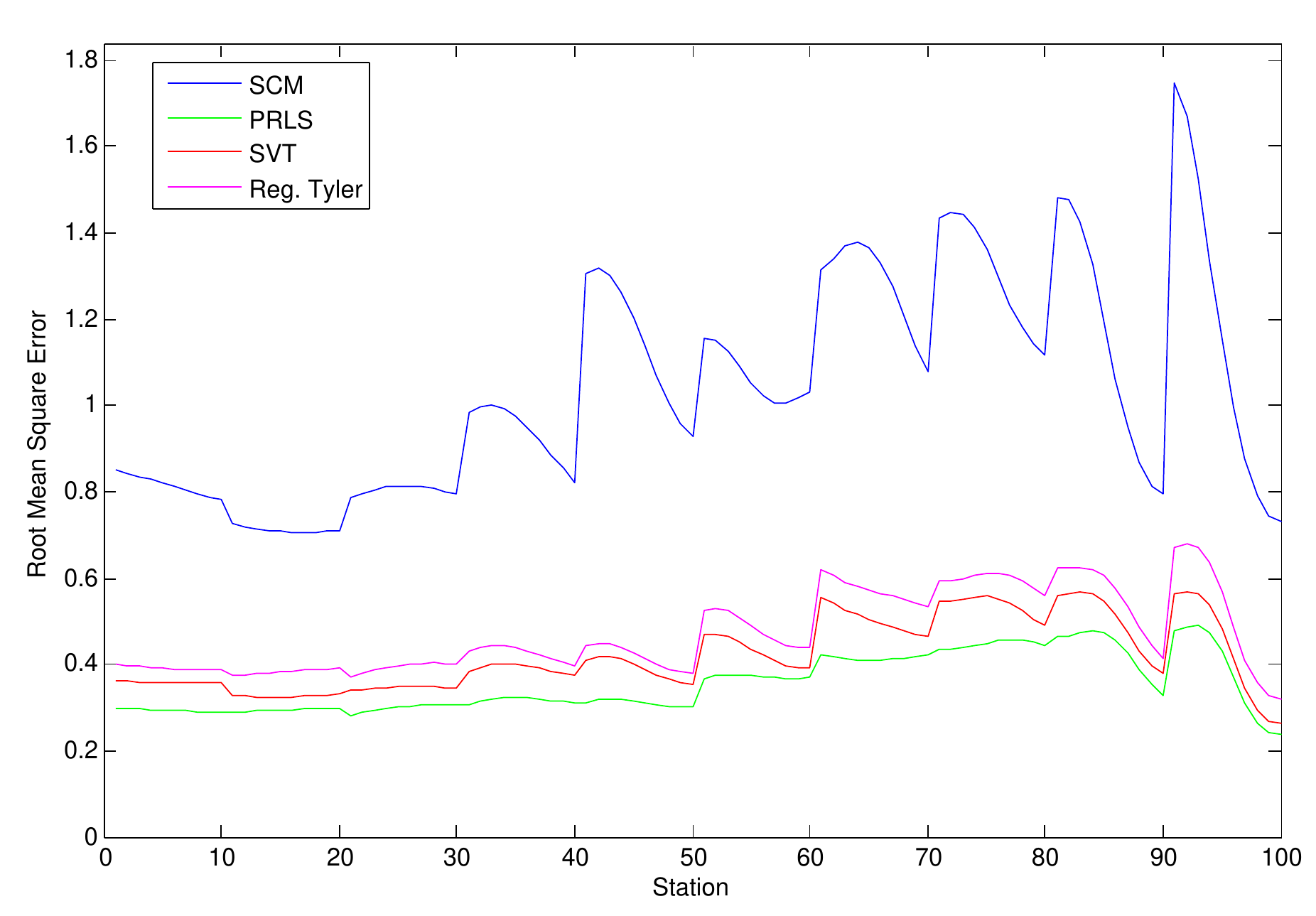}
	\caption{\label{fig:arctic_rmse_pred} NCEP wind speed data (Arctic Ocean): RMSE prediction performance across $q$ stations for linear estimators using SCM (blue) and PRLS (green). The estimators PRLS, SVT and regularized Tyler respectively achieve an average reduction in RMSE of $4.64$, $3.91$ and $3.41$ dB as compared to SCM (averaged across stations). }
\end{figure}

\begin{figure}[ht]
	\centering
		\includegraphics[width=1.00\textwidth]{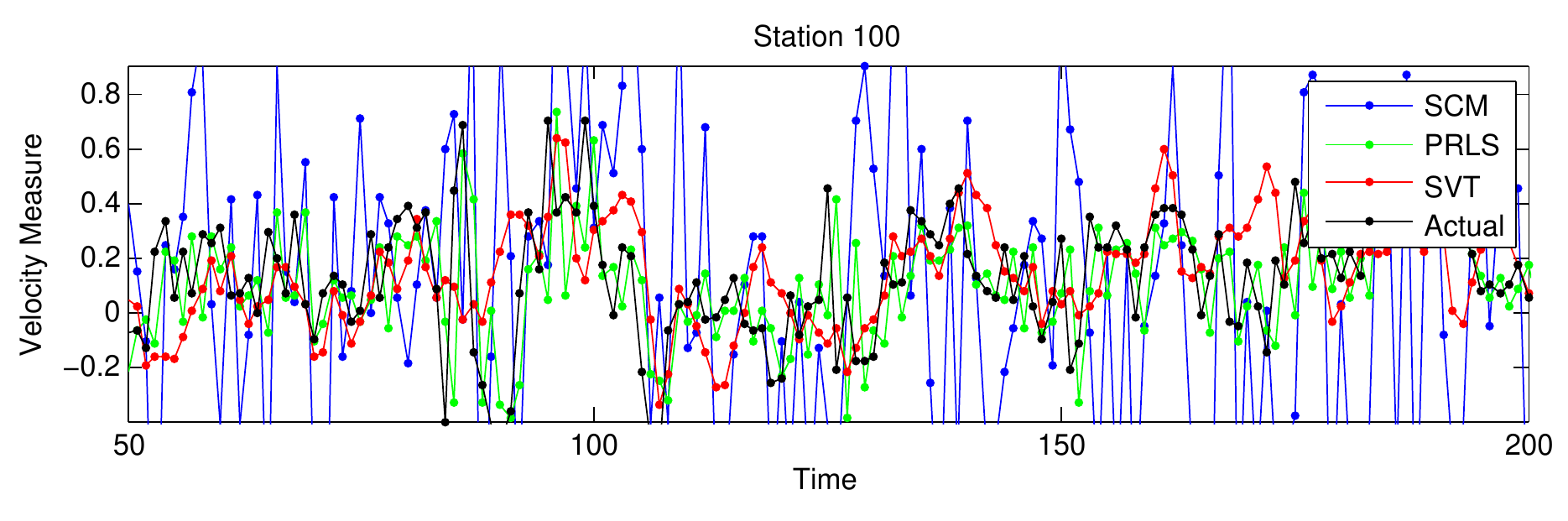}
	\caption{\label{fig:arctic_pred} NCEP wind speed data (Arctic Ocean): Prediction performance for linear estimators using SCM (blue), SVT (red) and PRLS (green) for a time interval of $150$ days. The actual (ground truth) wind speeds are shown in black. PRLS offers better tracking performance as compared to SCM and SVT. }
\end{figure}

Fig.~\ref{fig:arctic_rmse_pred} shows the root mean squared error (RMSE) prediction performance over the testing period of $1825$ days for the forecasts based on the standard SCM, PRLS, and regularized Tyler \cite{ChenWieselHero:2011}. The PRLS estimator was implemented using a regularization parameter $\lambda_n = C \nn \hat{\bS}_n \nn_2 \sqrt{\frac{p^2+q^2+\log(\max(p,q,n))}{n}}$ with $C=0.073$. The constant $C$ was chosen by optimizing the prediction RMSE on the training set over a range of regularization parameters $\lambda$ parameterized by $C$ (as in Irish wind speed data set). The SVT estimator proposed by Lounici \cite{Lounici} was implemented using a regularization parameter $\lambda = C \sqrt{\tr(\hat{S}_n) \nn\hat{S}_n\nn_2} \sqrt{\frac{\log(2pq)}{n}}$ with constant $C=0.47$ optimized in a similar manner. Fig.~\ref{fig:arctic_pred} shows a sample period of $150$ days. It is observed that SCM has unstable performance, while the Kronecker product estimator offers better tracking of the wind speeds.

\section{Conclusion}
We have introduced a framework for covariance estimation based on separation rank decompositions using a series of Kronecker product factors. We proposed a least-squares estimator in a permuted linear space with nuclear norm penalization, named PRLS. We established high dimensional consistency for PRLS with guaranteed rates of convergence. The analysis shows that for low separation rank covariance models, our proposed method outperforms the standard SCM estimator. For the class of block-Toeplitz matrices with exponentially decaying off-diagonal norms, we showed that the separation rank is small, and specialized our convergence bounds to this class. We also presented synthetic simulations that showed the benefits of our methods.

As a real world application we demonstrated the performance of the proposed Kronecker product-based estimator in wind speed prediction using an Irish wind speed dataset and a recent US NCEP dataset. Implementation of a standard covariance-based prediction scheme using our Kronecker product estimator achieved performance gains as compared to standard with respect to previously proposed covariance-based predictors.

There are several questions that remain open and are worthy of additional study. First, while the proposed penalized least squares Kronecker sum approximation yields a unique solution, the solution requires specification of the parameter $\lambda$, which specifies both the separation rank, and the amount of spectral shrinkage in the approximation. It would be worthwhile to investigate optimal or consistent methods of choosing this regularization parameter, e.g. using Stein's theory of unbiased risk minimization.  Second, while we have proven positive definiteness of the Kronecker sum approximation when the number of samples is greater than than the variable dimension in our experiments we have observed that positive definiteness is preserved more generally. Maximum likelihood estimation of Kronecker sum covariance and inverse covariance matrices is a worthwhile open problem. Finally, extensions of the low separation rank estimation method (PRLS) developed here to missing data follow naturally through the methodology of low rank covariance estimation studied in \cite{Lounici}.

\section*{Acknowledgement}
The research reported in this paper was supported in part by ARO grant W911NF-11-1-0391.

\appendices

\section{Proof of Theorem \ref{thm: symmetry_pd}}
\begin{IEEEproof}\\[0em]
1) \underline{Symmetry}\\[0em]
Recall the permuted version of the sample covariance $\hat{\bS}_n$, i.e., $\hat{\bR}_n=\mathcal{R}(\hat{\bS}_n)$. The SVD of $\hat{\bR}_n$ can be obtained as a solution to the minimum norm problem (Thm. 1 and Cor. 2 in \cite{VanLoanKP}, Sec. 3 in \cite{Tyrtyshnikov:2004}): 
\begin{equation} \label{eq: cov_match}
	\min_{\{\bA_k,\bB_k\}_k} \parallel \hat{\bS}_n - \sum_{k=1}^r \bA_k\otimes\bB_k \parallel_F^2
\end{equation}
subject to the orthogonality constraints $\tr(\bA_k^T \bA_l)=\tr(\bB_k^T \bB_l)=0$ for $k\neq l$. 
Since the Frobenius norm is invariant to permutations, we have the equivalent optimization problem:
\begin{equation} \label{eq: cov_match_svd}
	\min_{\{\bu_k,\bv_k\}_k} \parallel \hat{\bR}_n - \sum_{k=1}^r \sigma_k \bu_k\bv_k^T \parallel_F^2
\end{equation}
subject to the orthonormality conditions $\bu_k^T\bu_l=\bv_k^T\bv_l=1$ for $k=l$ and $0$ if $k\neq l$. The correspondence of (\ref{eq: cov_match}) with (\ref{eq: cov_match_svd}) is given by the mapping $\bu_k=\vec(\bA_k)$ and $\bv_k=\sigma_k \vec(\bB_k)$. The SVD of $\hat{\bR}_n$ can be written in matrix form as $\bU \bSigma \bV^T$.

We next show that the symmetry of $\hat{\bS}_n$ implies that the PRLS solution is symmetric by showing that the reshaped singular vectors $\bu_k$ and $\bv_k$ correspond to symmetric matrices. From the SVD definition \cite{MatrixComputations}, the right singular vectors $\bv_k$ are eigenvectors of $\bM_n = \hat{\bR}_n^T \hat{\bR}_n$ and thus satisfy the eigenrelation:
\begin{equation} \label{eq:eig}
	\bM_n \bv_k = \sigma_k^2 \bv_k
\end{equation}
where $\sigma_k = [\bSigma]_{k,k}$. Expressing (\ref{eq:eig}) in terms of the permutation operator $\mathcal{R}$, we obtain:
\begin{equation} \label{eq:eig_mat}
	\sum_{i,j=1}^p \left<\bv_k,\vec(\hat{\bS}_n(i,j))\right> \vec(\hat{\bS}_n(i,j)) = \sigma_k^2 \bv_k
\end{equation}
Define the $q\times q$ matrix $\bV_k$ such that $\bv_k=\vec(\bV_k)$. Rewriting (\ref{eq:eig_mat}) by reshaping vectors into matrices, we have after some algebra:
\begin{align}
	\sigma_k^2 \bV_k &= \sum_{i,j=1}^p \tr(\bV_k^T \hat{\bS}_n(i,j)) \hat{\bS}_n(i,j) \nonumber \\
		&= \underbrace{\sum_{i=1}^p \tr(\bV_k^T \hat{\bS}_n(i,i)) \hat{\bS}_n(i,i)}_{\bK_1} \nonumber \\
		&\quad + \underbrace{\sum_{i<j} \tr(\bV_k^T\hat{\bS}_n(i,j)) (\hat{\bS}_n(i,j)+\hat{\bS}_n(j,i))}_{\bK_2} \nonumber \\
		&\quad + \underbrace{\sum_{i<j} \tr(\bV_k^T(\hat{\bS}_n(j,i)-\hat{\bS}_n(i,j))) \hat{\bS}_n(j,i) }_{\bE} \label{eq:V_k}
\end{align}
Clearly, $\bK_1$ is symmetric since all submatrices $\hat{\bS}_n(i,i)$ are symmetric. Since $\hat{\bS}_n(j,i)=\hat{\bS}_n(i,j)^T$, it follows that $\bK_2$ is also symmetric. To finish the proof, we show $\bE=0$. Define the set
\begin{align*}
	\mathcal{L} &= \Big\{(i,j): i<j, \hat{\bS}_n(i,j) \neq 0, \hat{\bS}_n(i,j) \neq \hat{\bS}_n(j,i), \\
		&\qquad \hat{\bS}_n(i,j) \neq \hat{\bS}_n(i',j') \forall i'\neq i,j'\neq j \Big\}
\end{align*}
The set $\mathcal{L}$ is nonempty with probability 1 for any sample size. Let $l=\card(\mathcal{L})$. Then, we can rewrite:
\begin{equation} \label{eq:E}
	\bE = \sum_{(i,j) \in \mathcal{L}} \tr(\bV_k^T(\hat{\bS}_n(j,i)-\hat{\bS}_n(i,j))) \hat{\bS}_n(j,i)
\end{equation}
Since $\hat{\bS}_n(j,i)\neq 0$ with probability 1, $\bE=0$ iff $\tr(\bV_k^T(\hat{\bS}_n(j,i)-\hat{\bS}_n(i,j)))=0$ for all $i<j$. Using the properties of the trace operator, rewriting $\tr(\bV_k^T(\hat{\bS}_n(j,i)-\hat{\bS}_n(i,j)))=\tr((\bV_k^T-\bV_k)\hat{\bS}_n(j,i))$, we conclude from the decomposition $\sigma_k^2 \bV_k = \bK_1 + \bK_2 + \bE$ that $\bV_k=\bV_k^T$ if $\bE=0$. To finish the proof, we show that $\bE=0$ with probability 1. Taking the $\vec(\cdot)$ of (\ref{eq:E}), we conclude that $\bE=0$ is equivalent to 
\begin{equation} \label{eq:null}
	0 = \sum_{(i,j)\in \mathcal{L}} a_{i,j} \hat{\bS}_n(j,i)
\end{equation}
where $a_{i,j}=\tr((\bV_k^T-\bV_k)\hat{\bS}_n(j,i))$. The equation (\ref{eq:null}) can be rewritten as the linear equations:
\begin{equation} \label{eq:nulle}
	\bD \ba = \mathbf{0}
\end{equation}
where $\ba=\{a_{i,j}\}_{(i,j)\in \mathcal{L}} \in \RR^{l}$ and the columns of the $q^2 \times l$ matrix $\bD$ are given by $\bd_{i,j}=\vec(\hat{\bS}_n(j,i)) \in \RR^{q^2}$. Solutions of (\ref{eq:nulle}) are given by $\ba \in \Nul(\bD)$. Since the matrix $\bD$ is full-rank, $\ba=\mathbf{0}$ is the only solution of (\ref{eq:null}). This implies $\bE=0$, and therefore, $\bV_k=\bV_k^T$. Since $k$ is arbitrary, all reshaped right singular vectors of $\hat{\bR}_n$ are symmetric. A similar argument holds for all reshaped left singular vectors $\bu_k$. The proof is complete.



2) \underline{Positive Definiteness}\\[0em]
The sample covariance matrix $\hat{\bS}_n$ is positive definite with probability 1 if $n\geq pq$. First, consider the minimum norm problem (\ref{eq: cov_match}). The factors $\bA_k$ and $\bB_k$ are symmetric by part (1). If we show that a solution to (\ref{eq: cov_match}) has p.d. Kronecker factors, then the weighted sum with positive scalars is also p.d. and as a result, the PRLS solution given by $\hat{\bSigma}_n^\lambda=\sum_{k=1}^{r_0} \left( \sigma_k(\hat{\bR}_n)-\frac{\lambda}{2} \right)_+ \bU_k \otimes \bV_k$ is positive definite (see (\ref{eq:solution})).

Fix $l \in \{1,\dots,r_0\}$. We will show that in (\ref{eq: cov_match}) $\bA_k$ and $\bB_k$ can be restricted to be p.d. matrices. Define the eigendecompositions of $\bA_l$ and $\bB_l$:
\begin{align*}
	\bA_l &= \bPsi_l \bD_l \bPsi_l^T \\
	\bB_l &= \bXi_l \bLambda_l \bXi_l^T
\end{align*}
where $\{\bPsi_l\}_l, \{\bXi_l\}_l$ are sets of orthonormal matrices and $\bD_l, \bLambda_l$ are diagonal matrices. Let  $\bD_l=\diagg(d_l^1,\dots,d_l^p)$ and $\bLambda_l=\diagg(\lambda_l^1,\dots,\lambda_l^q)$. Set $\bQ_l = \bPsi_l \otimes \bXi_l$. Define $\bF_l = \bQ_l^T \hat{\bS}_n \bQ_l$. The objective function (\ref{eq: cov_match}) can be rewritten as:
\begin{align}
	\parallel \hat{\bS}_n &- \sum_{k=1}^r \bA_k \otimes \bB_k \parallel_F^2 \label{eq: obj_min_norm} \\
		&= \nn \bQ_l^T \left( \hat{\bS}_n - \sum_{k=1}^r \bA_k \otimes \bB_k \right) \bQ_l \nn_F^2 \nonumber \\
		&= \nn \bF_l - \sum_{k=1}^r \bQ_l^T(\bA_k \otimes \bB_k)\bQ_l  \nn_F^2 \nonumber \\
		&= \nn \underbrace{\bF_l -  \sum_{k\neq l} (\bPsi_l^T \bA_k \bPsi_l)\otimes(\bXi_l^T \bB_k \bXi_l) }_{\bM_l} - (\bPsi_l^T \bA_l \bPsi_l)\otimes(\bXi_l^T \bB_l \bXi_l) \nn_F^2 \nonumber \\
		&= \nn \bM_l - \bD_l \otimes \bLambda_l \nn_F^2 \nonumber \\
		&= \parallel \bM_l \nn_F^2 + \nn \bD_l \otimes \bLambda_l \parallel_F^2 - 2 \tr\left(\bF_l (\bD_l\otimes \bLambda_l)  \right) + 2 \sum_{k\neq l} \tr((\bPsi_l^T\bA_k\bPsi_l \otimes \bXi_l^T\bB_k\bXi_l)(\bD_l \otimes \bLambda_l)) \nonumber \\
		&= \nn \bM_l \nn_F^2 - \nn \bF_l \nn_F^2 + \nn \bF_l - \bD_l\otimes \bLambda_l \nn_F^2  + 2 \sum_{k\neq l} \tr(\bB_k \bB_l) \tr(\bA_k \bA_l) \nonumber \\
		&= \nn \bM_l \nn_F^2 - \nn \bF_l \nn_F^2 \nonumber \\
		&\quad + \nn \bF_l - \diagg(\bF_l) + \diagg(\bF_l) - \bD_l \otimes \bLambda_l \nn_F^2 \label{eq:orth1} \\
		&= \nn \bM_l \nn_F^2 - \nn \bF_l \nn_F^2 + \nn \bF_l - \diagg(\bF_l) \nn_F^2 + \nn\diagg(\bF_l) - \bD_l \otimes \bLambda_l \nn_F^2 \nonumber \\
		&\quad + 2 \tr\left( (\bF_l - \diagg(\bF_l)) (\diagg(\bF_l)-\bD_l \otimes \bLambda_l) \right) \nonumber \\
		&= \nn \bM_l \nn_F^2 - \nn \bF_l \nn_F^2 + \nn \bF_l - \diagg(\bF_l) \nn_F^2 \nonumber \\
		&\quad + \nn \diagg(\bF_l) - \bD_l\otimes \bLambda_l \nn_F^2 \label{eq:orth2}
\end{align}
where in equality (\ref{eq:orth1}) we used the orthogonality of Kronecker factors in the SVD. In equality (\ref{eq:orth2}), we used the fact that the matrices $\bF_l - \diagg(\bF_l)$ and $\diagg(\bF_l)-\bD_l \otimes \bLambda_l$ have disjoint support. We note that the term $\nn \bM_l \nn_F^2 - \nn \bF_l \nn_F^2 + \nn \bF_l - \diagg(\bF_l) \nn_F^2$ is independent of $\bD_l,\bLambda_l$. The positive definiteness of $\hat{\bS}_n$ implies that the diagonal elements of $\bF_l$ are all positive. Let $\diagg(\bF_l)=\diagg(\{f_{(i-1)q+j}\}_{i,j}) > 0$. Simple algebra yields:
\begin{align*}
	\nn &\diagg(\bF_l)- \bD_l\otimes \bLambda_l \nn_F^2 \\
		&= \sum_{i=1}^p \sum_{j=1}^q (f_{(i-1)q+j} - d_l^i \lambda_l^j)^2 = a_l + b_l
\end{align*}
where
\begin{align*}
		a_l &= \sum_{i=1}^p \sum_{j=1}^q (f_{(i-1)q+j} - |d_l^i| |\lambda_l^j|)^2 \\
		b_l &= 2 \sum_{i=1}^p \sum_{j=1}^q f_{(i-1)q+j} (|d_l^i||\lambda_l^j|-d_l^i \lambda_l^j)
\end{align*}
We note that the term $a_l$ is invariant to any sign changes of the eigenvalues $\{d_l^i,\lambda_l^j\}_{i,j}$ and the term $b_l$ is non-negative and equals zero iff $d_l^i,
\lambda_l^j$ have the same sign for all $i,j$. By contradiction, it follows that the eigenvalues $\{d_l^i\}_{i=1}^p$ and $\{\lambda_l^j\}_{j=1}^q$ must all have the same sign (if not, then the minimum norm is not achieved by $(\bA_l,\bB_l)$). Without loss of generality (since $\bA_l\otimes \bB_l = (-\bA_l)\otimes (-\bB_l)$, the signs can be assumed to be positive. We conclude that there exist p.d. matrices $(\bA_l,\bB_l)$ that achieve the minimum norm of (\ref{eq: obj_min_norm}). This holds for any $l$ so the proof is complete. 

\end{IEEEproof}

\section{Proof of Theorem \ref{thm: Frob_rate}}
\begin{IEEEproof}
The proof generalizes Thm. 1 in \cite{Lounici} to nonsquare matrices. A necessary and sufficient condition for the minimizer of (\ref{eq: prls_problem}) is that there exists a $\hat{\bV} \in \partial \nn \hat{\bR}^\lambda \nn_*$ such that:
\begin{equation} \label{eq:eq1}
	\left< 2 (\hat{\bR}^\lambda - \hat{\bR}_n) + \lambda \hat{\bV}, \hat{\bR}^\lambda - \bR \right> \leq 0
\end{equation}
for all $\bR$. From (\ref{eq:eq1}), we obtain for any $\bV \in \partial \nn \bR \nn_1$:
\begin{align}
	2&\left<\hat{\bR}^\lambda-\bR_0,\hat{\bR}^\lambda-\bR\right> + \lambda \left<\hat{\bV}-\bV,\hat{\bR}^\lambda-\bR\right> \nonumber \\
		&\leq -\lambda \left<\bV,\hat{\bR}^\lambda-\bR\right> + 2\left<\hat{\bR}_n-\bR_0,\hat{\bR}^\lambda-\bR\right> \label{eq:eq2}
\end{align}
The monotonicity of subdifferentials of convex functions implies:
\begin{equation} \label{eq:eq3}
	\left< \hat{\bV}-\bV,\hat{\bR}^\lambda-\bR \right> \geq 0
\end{equation}
From Example 2 in \cite{Watson}, we have the characterization of the subdifferential of a nuclear norm of a nonsquare matrix:
\begin{equation*}
	\partial \nn \bR \nn_* = \left\{ \sum_{j=1}^r \bu_j(\bR) \bv_j(\bR)^T  + \bP_U^{\perp} \bW \bP_V^{\perp} : \nn \bW \nn_2 \leq 1  \right\}
\end{equation*}
where $r = \rank(\bR)$, $U=\spann\{\bu_j\}$ and $V=\spann \{\bv_j\}$. Thus, for $\bR=\sum_{j=1}^r \sigma_j(\bR) \bu_j \bv_j^T$, $r=\rank(\bR)$, we can write:
\begin{equation}
	\bV = \sum_{j=1}^r \bu_j \bv_j^T + \bP_U^{\perp} \bW \bP_V^{\perp}
\end{equation}
where $\bW$ can be chosen such that $\nn\bW \nn_2 \leq1 $ and 
\begin{equation} \label{eq:eq4}
	\left< \bP_U^{\perp} \bW \bP_V^{\perp},\hat{\bR}^\lambda-\bR \right> = \nn \bP_U^{\perp} \hat{\bR}^\lambda \bP_V^{\perp}\nn_*
\end{equation}
Next, note the equality:
\begin{align}
	\nn\hat{\bR}^\lambda-\bR_0\nn_F^2 & + \nn\hat{\bR}^\lambda-\bR\nn_F^2 - \nn\bR-\bR_0\nn_F^2 \nonumber\\
		&= 2\left< \hat{\bR}^\lambda-\bR_0,\hat{\bR}^\lambda-\bR \right> \label{eq:eq5}
\end{align}
Using (\ref{eq:eq3}), (\ref{eq:eq4}) and (\ref{eq:eq5}) in (\ref{eq:eq2}), we obtain:
\begin{align}
	\nn & \hat{\bR}^\lambda-\bR_0\nn_F^2 + \nn\hat{\bR}^\lambda-\bR\nn_F^2 + \lambda \nn \bP_U^{\perp} \hat{\bR}^\lambda \bP_V^{\perp}\nn_* \nonumber \\
		& \leq \nn \bR-\bR_0\nn_F^2 + \lambda \left< \sum_{j=1}^r \bu_j \bv_j^T,-(\hat{\bR}^\lambda-\bR) \right> \nonumber \\
		& \quad + 2\left< \hat{\bR}_n-\bR_0,\hat{\bR}^\lambda-\bR \right> \label{eq:eq6}
\end{align}
From trace duality, we have:
\begin{align*}
	\Big<& \sum_{j=1}^r \bu_j \bv_j^T,-(\hat{\bR}^\lambda-\bR) \Big> \\
		&= \left<\bP_U \sum_{j=1}^r \bu_j \bv_j^T \bP_V, -(\hat{\bR}^\lambda-\bR) \right> \\
		&\leq \nn \sum_{j=1}^r \bu_j \bv_j^T\nn_2 \nn \bP_U^T (\hat{\bR}^\lambda-\bR) \bP_V^T \nn_* \\
		&= \nn \bP_U (\hat{\bR}^\lambda-\bR) \bP_V \nn_*
\end{align*}
where we used the symmetry of projection matrices. Using this bound in (\ref{eq:eq6}), we obtain:
\begin{align}
	\nn & \hat{\bR}^\lambda-\bR_0\nn_F^2 + \nn\hat{\bR}^\lambda-\bR\nn_F^2 + \lambda \nn \bP_U^{\perp} \hat{\bR}^\lambda \bP_V^{\perp}\nn_* \nonumber \\
		& \leq \nn \bR-\bR_0\nn_F^2 + \lambda \nn \bP_U (\hat{\bR}^\lambda-\bR) \bP_V \nn_* \nonumber \\
		& \quad + 2 \left< \bDelta_n,\hat{\bR}^\lambda-\bR \right> \label{eq:eq7}
\end{align}
where $\bDelta_n = \hat{\bR}_n-\bR_0$. Define the orthogonal projection of $\bR$ onto the outer product span of $U$ and $V$ as $\mathcal{P}_{U,V}(\bR)=\bR - \bP_U^{\perp} \bR \bP_V^{\perp}$. Then, we decompose:
\begin{align*}
	\left< \bDelta_n,\hat{\bR}^\lambda-\bR \right> &= \left<\bDelta_n,\mathcal{P}_{U,V}\left(\hat{\bR}^\lambda-\bR\right)\right> \\
		&\quad + \left< \bDelta_n, \bP_U^{\perp} (\hat{\bR}^\lambda-\bR) \bP_V^{\perp} \right>
\end{align*}
By the Cauchy-Schwarz inequality and trace-duality:
\begin{align*}
	\nn \bP_U (\hat{\bR}^\lambda-\bR) \bP_V \nn_* &\leq \sqrt{\rank(\bR)} \nn \hat{\bR}^\lambda - \bR \nn_F \\
	|\left< \bDelta_n,\mathcal{P}_{U,V}(\hat{\bR}^\lambda-\bR) \right>| &\leq \nn\bDelta_n\nn_2 \nn \mathcal{P}_{U,V}(\hat{\bR}^\lambda-\bR) \nn_* \\
		&\leq \nn\bDelta_n\nn_2 \sqrt{2\rank(\bR)} \nn \hat{\bR}^\lambda-\bR \nn_F \\
	|\left<\bDelta_n, \bP_U^{\perp} (\hat{\bR}^\lambda-\bR) \bP_V^{\perp}\right>| &\leq \nn\bDelta_n\nn_2 \nn \bP_U^{\perp} \hat{\bR}^\lambda \bP_V^{\perp}\nn_*
\end{align*}
where we used $\bP_U^{\perp} \bR \bP_V^{\perp} = 0$. Using these bounds in (\ref{eq:eq7}), we further obtain:
\begin{align}
	\nn & \hat{\bR}^\lambda-\bR_0\nn_F^2 + \nn\hat{\bR}^\lambda-\bR\nn_F^2 + (\lambda-2\nn\bDelta_n\nn_2) \nn \bP_U^{\perp} \hat{\bR}^\lambda \bP_V^{\perp}\nn_* \nonumber \\
		& \leq \nn \bR-\bR_0\nn_F^2 + ((2\sqrt{2} \nn\bDelta_n\nn_2+\lambda)\sqrt{r})(\sqrt{\nn \hat{\bR}^\lambda-\bR\nn_F^2}) \label{eq:eq8}
\end{align}
Using the arithmetic-mean geometric-mean inequality in the RHS of (\ref{eq:eq8}) and the assumption $\lambda \geq 2 \nn\bDelta_n\nn_2$, we obtain:
\begin{equation*}
	\nn \hat{\bR}^\lambda-\bR_0 \nn_F^2 \leq \nn \bR-\bR_0 \nn_F^2 + \frac{\lambda^2 (1+\sqrt{2})^2}{4} r
\end{equation*}
This concludes the proof.
\end{IEEEproof}


\section{Lemma \ref{lemma: large_dev}}
\begin{lemma} (Concentration of Measure for Coupled Gaussian Chaos) \label{lemma: large_dev}
	Let $\bx=[x_1,\dots,x_{p^2}]^T \in \mathcal{S}_{p^2-1}$ and $\by=[y_1,\dots,y_{q^2}]^T \in \mathcal{S}_{q^2-1}$. In the SCM (\ref{eq:SCM}) assume that $\{\bz_t\}$ are i.i.d. multivariate normal $\bz_t \sim N(0,\bSigma_0)$. Recall $\bDelta_n$ in (\ref{eq:bDelta_n}). For all $\tau \geq 0$:
	\begin{equation}
		\PP(|\bx^T\bDelta_n \by| \geq \tau) \leq 2 \exp\left( \frac{-n \tau^2/2}{C_1 \nn \bSigma_0 \nn_2^2 + C_2 \nn \bSigma_0\nn_2 \tau } \right)
	\end{equation}
	where $C_1 = \frac{4 e}{\sqrt{6\pi}} \approx 2.5044$ and $C_2 = e \sqrt{2} \approx 3.8442$ are absolute constants.
\end{lemma}
\begin{IEEEproof}
This proof is based on concentration of measure for Gaussian matrices and is similar to proof techniques used in compressed sensing (see Appendix A in \cite{Rauhut}) and in estimation of matrix variate normal models (see Appendix C in \cite{Gemini:2013}). Note that by the definition of the reshaping permutation operator $\mathcal{R}(\cdot)$, we have:
\begin{equation*}
	\bDelta_n = \frac{1}{n} \sum_{t=1}^n \begin{bmatrix} \vec(\bz_t(1) \bz_t(1)^T )^T - \E[\vec(\bz_t(1) \bz_t(1)^T )^T]  \\ \vdots \\ \vec(\bz_t(p) \bz_t(p)^T )^T - \E[\vec(\bz_t(p) \bz_t(p)^T )^T] \end{bmatrix}
\end{equation*}
where $\bz_t(i) = [\bz_t]_{(i-1)q+1:i q}$ is the $i$th subvector of the $t$th observation $\bz_t$. Thus, we can write:
\begin{equation*}
	\bx^T \bDelta_n \by = \frac{1}{n} \sum_{t=1}^n \psi_t
\end{equation*}
where
\begin{align}
	\psi_t &= \sum_{i,j=1}^p \sum_{k,l=1}^q \bX_{i,j} \bY_{k,l} \nonumber \\
		&\times ([\bz_t]_{(i-1)q + k}[\bz_t]_{(j-1)q + l}-\E[[\bz_t]_{(i-1)q + k}[\bz_t]_{(j-1)q + l}])  \label{eq:gaussian_chaos}
\end{align}
and $\bX \in \RR^{p\times p}$ and $\bY \in \RR^{q \times q}$ are reshaped versions of $\bx$ and $\by$. Defining $\bM = \bX \otimes \bY$, we can write (\ref{eq:gaussian_chaos}) as:
\begin{equation*}
	\psi_t = \bz_t^T \bM \bz_t - \E[\bz_t^T \bM \bz_t]
\end{equation*}
The statistic (\ref{eq:gaussian_chaos}) has the form of Gaussian chaos of order 2 \cite{Ledoux:1991}. Many of the random variables involved in the summation (\ref{eq:gaussian_chaos}) are correlated, which makes the analysis difficult. To simplify the concentration of measure derivation, using the joint Gaussian property of the data, we note that a stochastic equivalent of $\bz_t^T \bM \bz_t$ is $\bbeta_t^T \tilde{\bM} \bbeta_t$, where $\tilde{\bM} = \bSigma_0^{1/2} \bM \bSigma_0^{1/2}$, and $\bbeta_t \sim N(\mathbf{0},\bI_{pq})$ is a random vector with i.i.d. standard normal components. With this decoupling, we have:
\begin{align*}
	\E &|\psi_t|^2 = \E\left|\bbeta_t^T \tilde{\bM} \bbeta_t - \E[\bbeta_t^T \tilde{\bM} \bbeta_t] \right|^2 \\
		&= \E\left| \sum_{i_1\neq i_2} [\bbeta_t]_{i_1} [\bbeta_t]_{i_2} \tilde{\bM}_{i_1,i_2} + \sum_{i_1=1}^d ([\bbeta_t]_{i_1}^2-1) \tilde{\bM}_{i_1,i_1} \right|^2 \\
		&= \sum_{i_1\neq i_2} \sum_{i_1'\neq i_2'} \E[[\bbeta_t]_{i_1}[\bbeta_t]_{i_2}[\bbeta_t]_{i_1'}[\bbeta_t]_{i_2'}] \tilde{\bM}_{i_1,i_2} \tilde{\bM}_{i_1',i_2'} \\
		&\quad + \sum_{i_1} \sum_{i_1'} \E[([\bbeta_t]_{i_1}^2-1)([\bbeta_t]_{i_1'}^2-1)] \tilde{\bM}_{i_1,i_1} \tilde{\bM}_{i_1',i_1'} \\
		&= \sum_{i_1\neq i_2} \tilde{\bM}_{i_1,i_2}^2 + 2 \sum_{i_1} \tilde{\bM}_{i_1,i_1}^2 \\
		&= \nn\tilde{\bM}\nn_F^2 + \nn \diagg(\tilde{\bM}) \nn_F^2\\
		&\leq 2 \nn \tilde{\bM} \nn_F^2 \leq 2 \nn\bSigma_0\nn_2^2 \nn \bM \nn_F^2 = 2 \nn \bSigma_0 \nn_2^2
\end{align*}
where in the last step we used $\nn \bM \nn_F = \nn \bX \nn_F \nn \bY \nn_F = 1$.

Using a well known moment bound on Gaussian chaos (see p. 65 in \cite{Ledoux:1991}) and Stirling's formula, it can be shown (see, for example, Appendix A in \cite{Rauhut}) that for all $m\geq 3$:
\begin{equation}
	\E |\psi_t|^m \leq m! W^{m-2} v_t/2
\end{equation}
where
\begin{align*}
	W &= e \sqrt{\E|\psi_t|^2} \leq e\sqrt{2} \nn \bSigma_0 \nn_2 \\
	v_t &= \frac{2 e}{ \sqrt{6\pi} } \E|\psi_t|^2 \leq \frac{4e}{\sqrt{6\pi}} \nn \bSigma_0 \nn_2^2
\end{align*}
From Bernstein's inequality (see Thm. 1.1 in \cite{Rauhut}), we obtain:
\begin{align*}
	\PP\left(\left|\frac{1}{n} \sum_{t=1}^n \psi_t\right| \geq \tau\right) &\leq 2 \exp\left( \frac{-n^2 \tau^2/2}{ n v_1 + W n \tau} \right) \\
		&\leq 2 \exp\left( \frac{-n \tau^2/2}{ C_1 \nn\bSigma_0\nn_2^2 + C_2 \nn\bSigma_0\nn_2 \tau} \right)
\end{align*}
This concludes the proof.

\end{IEEEproof}

\section{Proof of Theorem \ref{thm: Operator_rate}}
\begin{IEEEproof}
Let $\mathcal{N}(\mathcal{S}_{d'-1},\epsilon')$ denote an $\epsilon'$-net on the $d'$-dimensional sphere $\mathcal{S}_{d'-1}$. Let $\bx_1 \in \mathcal{S}_{p^2-1} $ and $\by_1 \in \mathcal{S}_{q^2-1}$ be such that $|\bx_1^T\bDelta_n \by_1| = \nn \bDelta_n \nn_2$. By the definition of $\epsilon'$-net, there exists $\bx_2 \in \mathcal{N}(\mathcal{S}_{p^2-1},\epsilon')$ and $\by_2\in \mathcal{N}(\mathcal{S}_{q^2-1},\epsilon')$ such that $\nn\bx_1-\bx_2\nn_2\leq \epsilon'$ and $\nn \by_1-\by_2 \nn_2 \leq \epsilon'$. Then, by the Cauchy-Schwarz inequality:
\begin{align*}
	|&\bx_1^T\bDelta_n \by_1| - |\bx_2^T \bDelta_n \by_2| \leq |\bx_1^T\bDelta_n \by_1-\bx_2^T \bDelta_n \by_2| \\
		&= |\bx_1^T\bDelta_n (\by_1 - \by_2)> + (\bx_1 - \bx_2)^T\bDelta_n \by_2>| \\
		&\leq 2 \epsilon' \nn \bDelta_n \nn_2
\end{align*}
Since $\nn\bDelta_n \nn_2=|\bx_1^T \bDelta_n \by_1|$, this implies:
\begin{align*}
	\nn &\bDelta_n\nn_2 (1-2\epsilon')  \\
		&\leq \max\Big\{|\bx_2^T \bDelta_n \by_2|: \bx_2\in \mathcal{N}(\mathcal{S}_{p^2-1},\epsilon'), \by_2\in \mathcal{N}(\mathcal{S}_{q^2-1},\epsilon'), \\
		&\qquad \qquad \nn\bx_1-\bx_2\nn_2 \leq \epsilon', \nn\by_1-\by_2\nn_2 \leq \epsilon'\Big\} \\
		&\leq \max\left\{|\bx^T\bDelta_n \by|: \bx\in \mathcal{N}(S^{p^2-1},\epsilon'), \by\in \mathcal{N}(S^{q^2-1},\epsilon')\right\}
\end{align*}
As a result,
\begin{equation} \label{eq:op_norm_net}
	\nn \bDelta_n \nn_2 \leq (1-2\epsilon')^{-1} \max_{\bx \in \mathcal{N}(\mathcal{S}_{p^2-1},\epsilon'), \by \in \mathcal{N}(\mathcal{S}_{q^2-1},\epsilon') } |\bx^T \bDelta_n \by|
\end{equation}
From Lemma 5.2 in \cite{Vershynin}, we have the bound on the cardinality of the $\epsilon'$-net:
\begin{equation} \label{eq:card_bnd_sphere}
	\card(\mathcal{N}(\mathcal{S}_{d'-1},\epsilon')) \leq \left(1 + \frac{2}{\epsilon'}\right)^{d'}.
\end{equation}
From (\ref{eq:op_norm_net}), (\ref{eq:card_bnd_sphere}) and the union bound:
\begin{align*}
	\PP(& \nn \bDelta_n \nn_2 \geq \epsilon) \\
	  &\leq \PP\left(\max_{\bx \in \mathcal{N}(\mathcal{S}_{p^2-1},\epsilon'), \by \in \mathcal{N}(\mathcal{S}_{q^2-1},\epsilon')} |\bx^T\bDelta_n \by| \geq \epsilon (1-2\epsilon')\right) \\
		&\leq \PP\left(\bigcup_{\bx \in \mathcal{N}(\mathcal{S}_{p^2-1},\epsilon'), \by \in \mathcal{N}(\mathcal{S}_{q^2-1},\epsilon')} |\bx^T \bDelta_n \by| \geq \epsilon(1-2\epsilon') \right) \\
		&\leq \card(\mathcal{N}(\mathcal{S}_{p^2-1},\epsilon')) \card(\mathcal{N}(\mathcal{S}_{q^2-1},\epsilon')) \\
		&\quad \times \max_{\bx \in \mathcal{N}(\mathcal{S}_{p^2-1},\epsilon'), \by \in \mathcal{N}(\mathcal{S}_{q^2-1},\epsilon')} \PP(|\bx^T \bDelta_n \by|\geq \epsilon(1-2\epsilon')) \\
		&\leq \left(1 + \frac{2}{\epsilon'}\right)^{p^2+q^2} \PP\left(|\bx^T \bDelta_n \by| \geq \epsilon(1-2\epsilon')\right)
\end{align*}
Using Lemma \ref{lemma: large_dev}, we further obtain:
\begin{align}
	\PP(& \nn \bDelta_n \nn_2 \geq \epsilon) \nonumber \\
		&\leq 2 \left(1 + \frac{2}{\epsilon'}\right)^{p^2+q^2} \exp\left( \frac{-n\epsilon^2 (1-2\epsilon')^2/2}{C_1 \nn\bSigma_0\nn_2^2 + C_2\nn\bSigma_0 \nn_2 \epsilon (1-2\epsilon')} \right) \label{eq:bnd_gauss_tail}
\end{align}
We finish the proof by considering the two separate regimes. First, let us consider the Gaussian tail regime which occurs when $\epsilon \leq \frac{C_1 \nn \bSigma_0\nn_2}{C_2 (1-2\epsilon')}$. For this regime, the bound (\ref{eq:bnd_gauss_tail}) can be relaxed to:
\begin{align}
	\PP(& \nn \bDelta_n \nn_2 \geq \epsilon) \nonumber \\
		&\leq 2 \left(1 + \frac{2}{\epsilon'}\right)^{p^2+q^2} \exp\left( \frac{-n\epsilon^2 (1-2\epsilon')^2/2}{2C_1 \nn\bSigma_0\nn_2^2} \right) \label{eq:bnd_gauss_tail2}
\end{align}
Let us choose: 
\begin{equation*}
	\epsilon = \frac{t \nn \bSigma_0\nn_2}{1-2\epsilon'} \sqrt{\frac{p^2 + q^2 + \log M}{n}}
\end{equation*}
Then, from (\ref{eq:bnd_gauss_tail2}), we have:
\begin{align*}
	\PP &\left(\nn\bDelta_n \nn_2 \geq \frac{t \nn \bSigma_0\nn_2}{1-2\epsilon'} \sqrt{\frac{p^2 + q^2 + \log M}{n}} \right) \\
		&\leq 2 \left(1+\frac{2}{\epsilon'}\right)^{p^2+q^2} \exp\left( \frac{-t^2 (p^2 + q^2 + \log M)}{4C_1} \right) \\
		&\leq 2 \left( \left(1+\frac{2}{\epsilon'} \right) e^{-t^2/(4C_1)} \right)^{p^2+q^2} M^{-t^2/(4C_1)} \\
		&\leq 2 M^{-t^2/(4C_1)}
\end{align*}
This concludes the bound for the Gaussian tail regime. The exponential tail regime follows by similar arguments. Assuming $\epsilon \geq \frac{C_1 \nn\bSigma_0\nn_2}{C_2 (1-2\epsilon')}$, and setting $\epsilon=\frac{t\nn\bSigma_0\nn_2}{1-2\epsilon'} \frac{p^2+q^2+\log M}{n}$, we obtain from (\ref{eq:bnd_gauss_tail}): 
\begin{align*}
	\PP &\left(\nn\bDelta_n \nn_2 \geq \frac{t \nn \bSigma_0\nn_2}{1-2\epsilon'} \frac{p^2 + q^2 + \log M}{n} \right) \\
		&\leq 2 \left(1+\frac{2}{\epsilon'}\right)^{p^2+q^2} \exp\left( \frac{-t (p^2 + q^2 + \log M)}{4C_2} \right) \\
		&\leq 2 \left( \left( 1+\frac{2}{\epsilon'} \right) e^{-t/(4C_2)} \right)^{p^2+q^2} M^{-t/(4C_2)} \\
		&\leq 2 M^{-t/(4C_2)}
\end{align*}
where we used the assumption $t\geq 4C_2 \ln(1 + \frac{2}{\epsilon'})$. The proof is completed by combining both regimes and letting $C_0=\nn\bSigma_0\nn_2$ and noting that $t>1$, along with $\frac{t C_2}{C_1}>1$.

\end{IEEEproof}

\section{Proof of Theorem \ref{thm: PRLS_Frob_rate}}
\begin{IEEEproof}
Define the event
\begin{equation*}
	\mathcal{E}_r = \left\{\nn \hat{\bR}_n^\lambda-\bR_0 \nn_F^2 > \inf_{\bR:\rank(\bR)\leq r}\nn \bR-\bR_0\nn_F^2 + \frac{(1+\sqrt{2})^2}{4} \lambda_n^2 r \right\}
\end{equation*}
where $\lambda_n$ is chosen as in the statement of the theorem.

Theorem \ref{thm: Frob_rate} implies that on the event $\lambda_n \geq 2 \nn \bDelta_n \nn_2$, with probability 1, we have for any $1\leq r\leq r_0$:
\begin{equation*}
	\nn \hat{\bR}_n^\lambda - \bR_0 \nn_F^2 \leq \inf_{\bR: \rank(\bR)\leq r} \nn \bR-\bR_0 \nn_F^2 + \frac{(1+\sqrt{2})^2}{4} \lambda_n^2 r
\end{equation*}
Using this and Theorem \ref{thm: Operator_rate}, we obtain:
\begin{align*}
	\P\left( \mathcal{E}_r\right) &= \P\left(\mathcal{E}_r \cap \{\lambda_n \geq 2 \nn \bDelta_n \nn_2\}\right) + \P\left(\mathcal{E}_r \cap \{\lambda_n < 2 \nn \bDelta_n \nn_2\}\right) \\
		&\leq \cancelto{0}{\P(\mathcal{E}_r|\lambda_n \geq 2 \nn \bDelta_n\nn_2)}\P(\lambda_n \geq 2 \nn \bDelta_n\nn_2) \\
		&\qquad + \P\left(\lambda_n<2\nn\bDelta_n \nn_2\right) \\
		&=\P\Big(\nn\bDelta_n\nn_2 > \frac{C_0 t}{1-2\epsilon'} \\
		&\quad	\times \max\left\{\frac{p^2+q^2+\log M}{n}, \sqrt{\frac{p^2+q^2+\log M}{n}} \right\} \Big) \\
		&\leq 2 M^{-t/4C}
\end{align*}
This concludes the proof.

\end{IEEEproof}

\section{Proof of Lemma \ref{lemma: variational_bnd}}
\begin{IEEEproof}
From the min-max theorem of Courant-Fischer-Weyl \cite{HornJohnson}:
\begin{align*}
	\sigma_{k+1}^2(\bR) &= \lambda_{k+1}(\bR\bR^T) \\
		&= \min_{\mathcal{V}: \dim(\mathcal{V}^\perp)\leq k} \max_{\nn \bv \nn_2=1, \bv \in \mathcal{V}} \left< \bR\bR^T\bv,\bv \right>
\end{align*}
Define the set
\begin{equation*}
	\mathcal{V}_k = \{\bv\in \RR^{p^2}: \nn \bv \nn_2=1, \bv \perp \Col(\bR \bP_k \bR^T) \} \subset S^{p^2-1}.
\end{equation*}
Choosing $\mathcal{V}=\Col(\bR \bP_k \bR^T)^\perp$, we have the upper bound:
\begin{equation*}
	\sigma_{k+1}^2(\bR) \leq \max_{\bv \in \mathcal{V}_k} \left< \bR\bR^T \bv,\bv \right>
\end{equation*}
Using the definition of $\mathcal{V}_k$ and the orthogonality principle, we have:
\begin{align*}
	\left< \bR\bR^T \bv,\bv \right> &= \left< \bR(\bI-\bP_k)\bR^T \bv,\bv \right> \\
		&= \left< (\bI-\bP_k) \bR^T \bv,\bR^T \bv \right> \\
		&= \left< (\bI-\bP_k) \bR^T \bv,(\bI-\bP_k) \bR^T \bv \right> \\
		&= \nn (\bI-\bP_k) \bR^T \bv \nn_2^2
\end{align*}
Using this equality and the definition of the spectral norm \cite{HornJohnson}:
\begin{align*}
	\sigma_{k+1}^2(\bR) &\leq \max_{\bv \in \mathcal{V}_k} \nn (\bI-\bP_k) \bR^T \bv \nn_2^2 \\
		&\leq \max_{\bv \in S_{p^2-1}} \nn (\bI-\bP_k) \bR^T \bv \nn_2^2 \\
		&= \nn (\bI-\bP_k) \bR^T \nn_2^2
\end{align*}
Equality follows when choosing $\bP_k=\bV_k\bV_k^T$. This is seen by writing $\bI=\bV\bV^T$ and using the definition of the spectral norm and the sorting of the singular values. The proof is complete.
\end{IEEEproof}

\section{Proof of Theorem \ref{thm: approx_err_blk_toep}}
\begin{IEEEproof}
Note that $(\lambda,\bu)$ is an eigenvalue-eigenvector pair of the square symmetric matrix $\bR_0^T\bR_0$ if:
\begin{equation} \label{eq:eigen}
	\sum_{i,j} \vec(\bSigma_0(i,j)) \left<\bu,\vec(\bSigma_0(i,j))\right> = \lambda \bu
\end{equation}
So for $\lambda>0$, the eigenvector $\bu$ must lie in the span of the vectorized submatrices $\{\vec(\bSigma_0(i,j))\}_{i,j}$. Motivated by this result, we use the Gram-Schmidt procedure to construct a basis that incrementally spans more and more of the subspace $\spann(\{\vec(\bSigma_0(i,j))\}_{i,j})$. For the special case of the block-Toeplitz matrix, we have:
\begin{equation*}
	\spann(\{\vec(\bSigma_0(i,j))\}_{i,j}) = \spann(\{\vec(\bSigma(\tau))\}_{\tau=-N}^N)
\end{equation*}
where the mapping is given by $\bSigma_0(i,j)=\bSigma(j-i)$. Note that $\bSigma(-\tau)=\bSigma(\tau)^T$.

For simplicity, consider the case $k=2k'+1$ for some $k' \geq 0$. From Lemma \ref{lemma: variational_bnd}, we are free to choose an orthonormal basis set $\{\bv_1,\dots,\bv_k\}$ and form the projection matrix $\bP_k=\bV_k\bV_k^T$, where the columns of $\bV_k$ are the vectors $\{\bv_j\}$. We form the orthonormal basis using the Gram-Schmidt procedure \cite{HornJohnson}:
\begin{align*}
	\tilde{\bv}_0 &= \vec(\bSigma(0)), \\
		&\quad \bv_0 = \frac{\tilde{\bv}_0}{\nn \tilde{\bv}_0 \nn_2} \\
	\tilde{\bv}_1 &= \vec(\bSigma(1)) - \frac{\left<\vec(\bSigma(1)),\tilde{\bv}_0\right>}{\nn \tilde{\bv}_0 \nn_2^2} \tilde{\bv}_0, \\
		&\quad \bv_1 = \frac{\tilde{\bv}_1}{\nn \tilde{\bv}_1 \nn_2} \\
	\tilde{\bv}_{-1} &= \vec(\bSigma(-1)) - \frac{\left<\vec(\bSigma(-1)),\tilde{\bv}_0\right>}{\nn \tilde{\bv}_0 \nn_2^2} \tilde{\bv}_0 \\
		&- \frac{\left<\vec(\bSigma(-1)),\tilde{\bv}_1\right>}{\nn \tilde{\bv}_1 \nn_2^2} \tilde{\bv}_1, \\
		&\quad \bv_{-1} = \frac{\tilde{\bv}_{-1}}{\nn \tilde{\bv}_{-1} \nn_2} \\
		& etc.
\end{align*}
With this choice of orthonormal basis, it follows that for every $k=2k'+1$, we have the orthogonal projector:
\begin{equation*}
	\bP_k = \bv_0 \bv_0^T + \sum_{l=1}^{k'} (\bv_l\bv_l^T + \bv_{-l}\bv_{-l}^T)
\end{equation*}
This corresponds to a variant of a sequence of Householder transformations \cite{HornJohnson}. Using Lemma \ref{lemma: variational_bnd}:
\begin{align}
	\sigma_{k+1}^2(\bR_0) &\leq \nn \bR_0(\bI - \bP_k) \nn_2^2 \nonumber \\
		&\leq \nn \bR_0 - \bR_0\bP_k \nn_F^2 \nonumber\\
		&\leq p \sum_{l=k'+1}^{N} \nn \bSigma(l) \nn_F^2 + \nn \bSigma(-l) \nn_F^2 \label{eq:row_subtract} \\
		&\leq 2C' pq \sum_{l=k'+1}^N u^{2l} \nonumber\\
		&\leq 2C' pq \frac{u^{2k'+2}}{1-u^2} \nonumber\\
		&\leq 2C' pq \frac{u^k}{1-u^2} \nonumber
\end{align}
where we used Lemma \ref{lemma: LS_bound} to obtain (\ref{eq:row_subtract}).
To finish the proof, using the bound above and (\ref{eq:approx_error}):
\begin{align*}
	\inf_{\bR:\rank(\bR)\leq r} \nn \bR-\bR_0 \nn_F^2 &= \sum_{k=r}^{r_0-1} \sigma_{k+1}^2(\bR_0) \\
		&\leq \frac{2 C' pq}{1-u^2} \sum_{k=r}^{r_0-1} u^k  \\
		&\leq 2 C' pq \frac{u^r}{(1-u)^2}
\end{align*}
The proof is complete.
\end{IEEEproof}

\section{Lemma \ref{lemma: LS_bound}}

\begin{lemma} \label{lemma: LS_bound}
Consider the notation and setting of proof of Thm. \ref{thm: approx_err_blk_toep}. Then, for the projection matrix $\bP_k$ chosen, we have for $k = 2 k' + 1, k' \geq 1$:
\begin{equation*}
	\sigma_{k+1}^2(\bR_0) \leq \nn \bR_0 - \bR_0\bP_k \nn_F^2 \leq p \sum_{l=k'+1}^{N} \nn \bSigma(l) \nn_F^2 + \nn \bSigma(-l) \nn_F^2
\end{equation*}
\end{lemma}
\begin{IEEEproof}
	To illustrate the row-subtraction technique, we consider the simplified scenario $k'=1$. The proof can be easily generalized to all $k'\geq 1$. Without loss of generality, we write the permuted covariance
	\begin{equation} \label{eq:Sigma_0}
		\bSigma_0 = \begin{bmatrix} \bSigma(0) & \bSigma(1) \\ \bSigma(-1) & \bSigma(0) \end{bmatrix}
	\end{equation}
	as:
	\begin{equation*}
		\bR_0 = \mathcal{R}(\bSigma_0) = \begin{bmatrix} \vec(\bSigma(0))^T \\ \vec(\bSigma(1))^T \\ \vec(\bSigma(-1))^T \\  \vec(\bSigma(0))^T \end{bmatrix}
	\end{equation*}
	Using the Gram-Schmidt submatrix basis construction of the proof of Thm. \ref{thm: approx_err_blk_toep}, the sequence of projection matrices can be written as:
	\begin{align*}
		\bP_1 &= \bv_0\bv_0^T \\
		\bP_2 &= \bv_0\bv_0^T + \bv_1\bv_1^T \\
		\bP_3 &= \bv_0\bv_0^T + \bv_1\bv_1^T + \bv_{-1}\bv_{-1}^T
	\end{align*}
	where $\bv_i$ is the orthonormal basis constructed in the proof of Thm. \ref{thm: approx_err_blk_toep}. The singular value bound $\sigma_1^2(\bR_0) \leq \nn \bR_0 \nn_F^2 = 2\nn \bSigma(0) \nn_F^2 + \nn \bSigma(1) \nn_F^2 + \nn \bSigma(-1) \nn_F^2 $ is trivial \cite{HornJohnson}.
	
	For the second singular value, we want to prove the bound:
	\begin{equation} \label{eq:bnd_sval_2}
			\sigma_2^2(\bR_0) \leq \nn \bSigma(1) \nn_F^2 + \nn \bSigma(-1) \nn_F^2
	\end{equation}
	To show this, we use the variational bound of Lemma \ref{lemma: variational_bnd}:
	\begin{align*}
		\sigma_2^2(\bR_0) &\leq \nn \bR_0 - \bR_0 \bP_1 \nn_F^2 \\
			&=\left|\left| \begin{bmatrix} \vec(\bSigma(0))^T - \left<\vec(\bSigma(0)),\bv_0\right> \bv_0^T \\ \vec(\bSigma(1))^T - \left<\vec(\bSigma(1)),\bv_0\right>\bv_0^T \\ \vec(\bSigma(-1))^T-\left<\vec(\bSigma(-1)),\bv_0\right>\bv_0^T \\ \vec(\bSigma(0))^T - \left<\vec(\bSigma(0)),\bv_0\right> \bv_0^T \end{bmatrix} \right|\right|_F^2 \\
			&= \left|\left| \begin{bmatrix} \mathbf{0}^T \\ \vec(\bSigma(1))^T - \left<\vec(\bSigma(1)),\bv_0\right>\bv_0^T \\ \vec(\bSigma(-1))^T-\left<\vec(\bSigma(-1)),\bv_0\right>\bv_0^T \\ \mathbf{0}^T \end{bmatrix} \right|\right|_F^2 \\
			&= \nn \vec(\bSigma(1)) - \left<\vec(\bSigma(1)),\bv_0\right>\bv_0 \nn_2^2 \\
			&\quad + \nn \vec(\bSigma(-1))-\left<\vec(\bSigma(-1)),\bv_0\right>\bv_0 \nn_2^2 \\
			&\leq \nn \bSigma(1) \nn_F^2 + \nn \bSigma(-1) \nn_F^2
	\end{align*}
	where in the last step, we used the Pythagorean principle from least-squares theory \cite{Papoulis}-i.e.$\nn \bA - \frac{<\bA,\bB>}{\nn \bB \nn_F^2}\bB \nn_F^2 \leq \nn \bA \nn_F^2$ for any matrices $\bA,\bB$ of the same order.
	Next, we want to show
	\begin{equation} \label{eq:bnd_sval_3}
		\sigma_3^2(\bR_0) \leq \nn \bSigma(-1) \nn_F^2
	\end{equation}
	Define $\gamma(j) = \vec(\bSigma(j)) - \left<\vec(\bSigma(j)),\bv_0\right>\bv_0$. Using similar bounds and the above, after some algebra:
	\begin{align*}
		\sigma_3^2(\bR_0) &\leq \nn \bR_0 - \bR_0 \bP_2 \nn_F^2 \\
			&= \left|\left| \begin{bmatrix} \mathbf{0}^T  \\ \gamma(1)^T - \left<\vec(\bSigma(1)),\bv_1\right>\bv_1^T \\ \gamma(-1)^T - \left<\vec(\bSigma(-1)),\bv_1\right>\bv_1^T \\ \mathbf{0}^T \end{bmatrix} \right|\right|_F^2 \\
			&= \nn \vec(\bSigma(-1))^T - \left<\vec(\bSigma(-1)),\bv_0\right>\bv_0^T  \\
			&\quad - \left<\vec(\bSigma(-1)),\bv_1\right>\bv_1^T \nn_2^2 \\
			&= \nn \vec(\bSigma(-1))^T \nn_2^2 - |\left<\vec(\bSigma(-1)),\bv_0\right>|^2 \\
			&\quad - |\left<\vec(\bSigma(-1)),\bv_1\right>|^2 \\
			&\leq \nn \bSigma(-1) \nn_F^2
	\end{align*}
	where we observed that $\gamma(1) = \left<\vec(\bSigma(1)),\bv_1\right>\bv_1$ and used the Pythagorean principle again.
	
	Using $\bP_3$ and similar bounds, it follows that $\sigma_4^2(\bR_0)=0$, which makes sense since the separation rank of (\ref{eq:Sigma_0}) is at most 3. Generalizing to $k'\geq 1$ and noting that $\nn \bSigma_0 \nn_F^2 = p \nn\bSigma(0)\nn_F^2 + \sum_{l=1}^{p-1} (p-l) \nn \bSigma(l) \nn_F^2 + \nn \bSigma(l) \nn_F^2 \leq p \nn\bSigma(0)\nn_F^2 + p \sum_{l=1}^{p-1} \nn \bSigma(l) \nn_F^2 + \nn \bSigma(-l) \nn_F^2$, we conclude the proof.
\end{IEEEproof}

\bibliographystyle{IEEEtran}
\bibliography{myRefs}

\end{document}